\documentclass[aps,nofootinbib,twocolumn,preprintnumbers,superscriptaddress,floatfix,eqsecnum]{revtex4}

\usepackage{graphicx}
\usepackage{epsfig}
\usepackage{epstopdf}
\usepackage{rotating}
\usepackage[mathscr]{eucal}
\usepackage{amsmath,amssymb,bm,graphics}
\usepackage[pdftex,letterpaper=true]{hyperref}
\usepackage{yfonts}
\usepackage{slashed}

\newcommand\p{k}
\newcommand\omg{\gamma}
\newcommand\ipj{r}
\newcommand\pipj{r}

\begin{document}

\title
{Using Higher Moments of Fluctuations and their Ratios in the Search for the QCD Critical Point}

\author{Christiana~Athanasiou}
\affiliation{Department of Physics, Massachusetts Institute of Technology, Cambridge, MA 02139, USA}
\author{Krishna~Rajagopal}
\affiliation{Department of Physics, Massachusetts Institute of Technology, Cambridge, MA 02139, USA}
\author{Misha~Stephanov}
\affiliation{Department of Physics, University of Illinois, Chicago, Illinois 60607, USA}

\date{\today}

\begin{abstract} 
The QCD critical point can be found in heavy ion collision experiments via the non-monotonic behavior of many fluctuation observables as a function of the collision energy.  The event-by-event fluctuations of various particle multiplicities 
are enhanced in those collisions that freeze out near the critical point.  Higher, non-Gaussian, moments of the event-by-event distributions of such observables are particularly sensitive to critical fluctuations, since their magnitude depends on the critical correlation length to a high power.  We present quantitative estimates of the contribution of critical fluctuations to the third and fourth moments of the pion, proton and net proton multiplicities, 
as well as estimates of various measures of pion-proton correlations, all as a function of the same five non-universal parameters, one of which is the correlation length that parametrizes proximity to the critical point.  We show how to use nontrivial but parameter independent ratios among these more than a dozen fluctuation observables to discover the critical point. We also construct ratios that, if the critical point is found, can be used to overconstrain the values of the non-universal parameters.  
\end{abstract} 

\maketitle

\section{Introduction and Illustrative Results}









One of the main goals of heavy ion collision experiments is to map the phase diagram of QCD. The second-order critical point at which the first-order transition between hadron matter and quark-gluon plasma (QGP) ends is one of the distinctive features of the phase diagram. We currently do not have a systematic way of locating this point from first principles as model and lattice calculations face many challenges and much work still needs to be done in order to overcome them.  (For reviews, see Refs.~\cite{Stephanov:2004wx,Schmidt:2006us,Stephanov:2007fk,Koch:2008ia,Gupta:2009mu,Philipsen:2009dn,Schmidt:2009qq,Li:2010dy,Fukushima:2010bq}.)  
In the meantime, if the critical point is located in a region accessible to heavy-ion collision experiments, it can be discovered experimentally.
Experiments 
with this goal are underway and planned at the Relativistic Heavy Ion Collider (RHIC) at the Brookhaven National Laboratory (BNL) and at the Super Proton Synchrotron (SPS) at 
CERN in Geneva~\cite{Stephans:2008zz,Mohanty:2009vb,Schuster:2009ak,:2009vj}. It is therefore important to define, evaluate the utility of, and select
experimental observables that will allow us to locate the critical point, if it is located in an experimentally accessible region.

In heavy-ion collision experiments, the center of mass energy $\sqrt{s}$  is varied, thus changing the temperature and chemical potential of the produced matter and in this way scanning the phase diagram. The observed collective flow of the produced matter at RHIC strongly suggests the production of a strongly-coupled quark-gluon plasma~\cite{RHIC}. As the QGP expands and cools, it follows a path on the phase diagram 
that is characterized by approximately constant entropy density to baryon number density ratio until freeze-out, after when there are no further interactions that change the multiplicities of hadron species. When these particles are then detected, they give us information about the state of the matter at the freeze-out point. Therefore, in order to see the effects of the critical point on observables, one should try to get the freeze-out point as close to the critical point as possible by varying the collision center of mass energy, $\sqrt{s}$. 
Decreasing $\sqrt{s}$ decreases the entropy to baryon number ratio, and therefore corresponds to increasing the baryon chemical potential $\mu_B$ at freeze-out.

Present lattice calculations evade the fermion sign problem in different ways that all rely upon the smallness of $\mu_B/(3 T)$.  Although each is currently limited by systematic effects, and they do not give consistent guidance as to the location of the critical point, all present lattice calculations agree that it is  not found at $\mu_B<T$, where the calculations are most reliable~\cite{Fodor:2004nz,Allton:2005gk,Gavai:2008zr,deForcrand:2008vr}.  For this reason, experimental searches focus on collisions which freezeout with $\mu_B > 150$~MeV.  The upper extent of the experimentally accessible region of the phase diagram is determined by the largest freezeout $\mu_B$ at which collisions still have a high enough $\sqrt{s}$ that the matter they produce reaches temperatures in the transition region.

Upon scanning in $\sqrt{s}$ and thus in $\mu_B$, one should then be able to locate (or rule out the presence of) the critical point 
by using observables that are sensitive to the proximity of the freeze-out point to the critical 
point~\cite{Stephanov:1998dy,Stephanov:1999zu}. For example, 
for particles like pions and protons that interact with the critical mode, 
the fluctuations in the number of particles  in a given acceptance window
will increase near the critical point as the critical mode becomes massless and develops large long-wavelength correlations. As we vary $\sqrt{s}$, therefore, if the freeze-out point approaches the critical point, we would see an increase in the fluctuations in the number of those particles which interact with the critical mode.   These fluctuations would then decrease as we move away from the critical point. (This is true for any observables which are sensitive to the proximity of the critical point to the point where freeze-out occurs.) Hence, a characteristic signature of the critical point is the non-monotonic behavior of such variables, as a function 
of $\sqrt{s}$~\cite{Stephanov:1998dy,Stephanov:1999zu}. Another way to change the freeze-out point is by changing the size of the system by varying the centrality of the collisions, since larger systems freeze-out later and hence at somewhat smaller temperatures.

In this paper we describe how to use the increase in fluctuations of particle numbers
near the critical point as a probe to determine its location. The way one characterizes the fluctuations of an observable is by measuring it in each event in an ensemble of many events, and then measuring the variance and higher, non-Gaussian, moments of the event-by-event distribution of the observable.  
The contribution of the critical fluctuations to these moments is proportional to some positive power of $\xi$, the correlation length which, in the idealized thermodynamic limit, diverges at the critical point. In reality, $\xi$ reaches a maximum value at the critical point but does not diverge because as it cools the system spends only  a finite time in the vicinity of the critical point.  The system also has only a finite size, but it turns out that the finite time is a more stringent limitation on the growth of the correlation length~\cite{Stephanov:1999zu,Berdnikov:1999ph}. Estimates of the rate of growth of $\xi$ as the collision cools past the critical point (which take into account the phenomenon of critical slowing down) suggest that the maximal value of $\xi$ that can be reached is around $1.5-3$ fm \cite{Berdnikov:1999ph,Son:2004iv,Nonaka:2004pg}, compared to the natural $\sim 0.5$ fm away from the critical point. Higher moments depend on higher powers of $\xi$, making them more favorable in searching for the critical point \cite{Stephanov:2008qz}. In this paper we consider the second, third and fourth cumulants of particle multiplicity distributions for pions and protons. We also consider mixed pion-proton cumulants, again up to fourth order.

Our goal in this Introduction is to provide an illustrative example of one possible experimental outcome.  In Section I.A we define the observables that must be measured at each $\sqrt{s}$. In Section I.B we suppose that the critical point is located at $\mu_B=400$~MeV 
and then guess how the correlation length $\xi$ at freezeout will vary with the chemical potential $\mu_B$, and hence with $\sqrt{s}$, in a heavy ion collision program in which the beam energy is scanned.  In Section I.C we plot results for how seven of the observables that we define will vary with $\mu_B$, if the guess for $\xi(\mu_B)$ that we have made for illustrative purposes were to prove correct.  In Section II we provide the calculation of all the observables that we define, as a function of $\xi$, the proton and pion number densities, and four nonuniversal parameters that must ultimately be obtained from data. In Section III we construct ratios of observables that allow us to measure four combinations of $\xi$ and the four parameters.  And, we construct five ratios of observables which receive a contribution from critical fluctuations that is independent of $\xi$ and independent of all four currently poorly known parameters.  This means that we make robust predictions for these five ratios, predictions that could be used to provide a stringent  check on whether enhanced fluctuations discovered in some experimental data set are or are not due to critical fluctuations.  We close in Section IV with a discussion of remaining open questions.

We shall find that critical fluctuations can easily make contributions to the higher moments of the proton multiplicity distribution that are larger than those in a Poisson distribution by more than a factor of 100.  In Appendix A we convince ourselves that we can construct  a reasonable looking, but somewhat {\it ad hoc}, distribution whose higher moments are this large.  What we are able to calculate in Section II is moments of the distribution, not the distribution itself. In Appendix A we construct a toy model distribution that has moments comparable to those we calculate.  We also use this toy model to obtain a crude gauge of how our results would be modified by any effects that serve to limit the maximum proton multiplicity in a single event.  

In Appendix B we apply our calculation to determine the contribution of critical fluctuations to the third and fourth cumulants of the event-by-event distribution of the mean transverse momentum of the pions in an event.  We find that the critical contribution to these non-Gaussian cumulants are quite small, smaller even than the contributions of Bose-Einstein statistics.  For this reason, throughout the main text of the paper we focus entirely on number fluctuations, rather than transverse momentum fluctuations.

\subsection{Moments and cumulants of fluctuations}
\label{sec:cumulant}

We expect to see a peak in the Gaussian and non-Gaussian cumulants of particle multiplicity distributions near the critical point as we change $\sqrt{s}$. In this subsection, we describe how to calculate these higher cumulants from experimental data.  

Consider an ensemble of events in each of which we have measured the number of particles of two species,
which we shall denote $x$ and $y$.   The possibilities for $x$ and $y$ that we consider later include the number of pions $N_\pi$, the number of protons $N_p$, and the number of protons minus antiprotons  $N_{p-\bar p}\equiv N_p - N_{\bar p}$.     In each case, the number that is tallied should be the number of particles of the desired species near mid-rapidity in a specified window of rapidity.  This window in rapidity should be at least about one unit wide, in
order for our results to apply without significant acceptance
corrections~\cite{Stephanov:2001zj}.  Furthermore, the longitudinal expansion of the matter produced in the collision reduces correlations among particles separated by much more
than one unit in rapidity~\cite{Stephanov:2001zj}, making larger windows unnecessary.

We denote the average value of $x$ and $y$ over the whole ensemble of events by $\langle x\rangle$ and $\langle y\rangle$. Throughout this paper, we use single angle brackets to indicate the ensemble average of a quantity whose event-by-event distribution has been measured. And, we shall denote the deviation of $x$ and $y$ from their mean in a single event by
\begin{eqnarray}
\delta x &\equiv& x-\langle x \rangle\nonumber\\
\delta y &\equiv& y-\langle y \rangle
\end{eqnarray}
We now define the cumulants of the event-by-event distribution of a single observable, say $x$.  The second and third cumulants are given by
\begin{eqnarray}
 \label{eq:var}
\kappa_{2x} &\equiv& \langle\langle x^2\rangle\rangle \equiv \langle \, (\delta x)^2 \,\rangle\\
\label{eq:3rd-cumulant}
  \kappa_{3x} &\equiv& \langle\langle x^3\rangle\rangle \equiv \langle \, (\delta x)^3 \,\rangle\ ,
\end{eqnarray}
where we have introduced two equivalent notations for the cumulants.  The second cumulant $\kappa_{2x}$ is the variance of the distribution, while the skewness of the distribution is given by $\kappa_{3x}/\kappa_{2x}^{3/2}$.
%
%
%
The fourth cumulant is different from the
corresponding fourth moment:
\begin{equation}
\kappa_{4x} \equiv \langle \langle x^4 \rangle \rangle \equiv \langle \, (\delta x)^4 \, \rangle - 3\:\langle \, (\delta x)^2 \, \rangle^2 \ . \label{eq:4x}
\end{equation}
The kurtosis of the distribution is given by $\kappa_{4x}/\kappa_{2x}^2$.

The defining property of the cumulants is their additivity for
independent variables. For example, if $a$ and $b$ are two independent
random variables, then $\kappa_{i(a+b)}=\kappa_{ia}
+\kappa_{ib}$. This property is easily seen from the cumulant generating function
\begin{equation}
g(\mu) = \log \langle e^{\mu \,\delta x} \rangle\ ,
\end{equation}
which is manifestly additive.   The $n$'th cumulant of the $x$-distribution is 
given by
\begin{equation}
\kappa_{nx} = \left.\frac{\partial^n g(\mu)}{\partial \mu^n}\right|_{\mu=0}\ .
\end{equation}
Using the double bracket notation introduced above, $g(\mu)=\langle\langle e^{\mu x}\rangle\rangle$.
As a result of their additivity, cumulants of {\em extensive} variables,
such as $N_p$ or $N_\pi$, are all themselves extensive, meaning that they are proportional to the volume of the system $V$ in the thermodynamic limit. 

We shall also consider mixed cumulants, which generalize
the more familiar Gaussian measures of correlations to non-Gaussian measures. 
These are generated by
\begin{equation}
g(\mu,\nu)\equiv\sum_{n,m}\frac{\kappa_{nxmy}\,\mu^n\nu^m}{m!\,n!}
=\log\langle e^{\mu \, \delta x+\nu \, \delta y}\rangle\ ,
\end{equation}
and, for example, are given by
\begin{align} \label{eq:mixed1}
\kappa_{1x1y} &\equiv \langle\langle xy\rangle\rangle  =  \langle\, \delta x\,\delta y \,\rangle \ , \\
\kappa_{1x2y} &\equiv \langle\langle x y^2\rangle\rangle = \langle\, \delta x \, (\delta y)^2  \, \rangle\ , \\
\kappa_{2x2y} &\equiv \langle \langle x^2 y^2 \rangle \rangle\nonumber\\
&= \langle\, (\delta x)^2\, (\delta y)^2 \,\rangle - 2\langle \,\delta x \,\delta y\,\rangle^2 -  \langle \, (\delta x)^2 \, \rangle\: \langle \, (\delta y)^2 \, \rangle \ ,\\
\kappa_{1x3y} &\equiv \langle \langle x y^3 \rangle \rangle\nonumber\\
&=\langle \, \delta x \,(\delta y)^3\,\rangle  -3\:\langle \,\delta x\,\delta y\,\rangle\:\langle \,(\delta y)^2\,\rangle\ .  \label{eq:mixed4}
\end{align}
For two extensive variables $x$ and $y$ such mixed cumulants are also extensive, proportional to $V$.

We have described how to obtain the cumulants $\kappa_{ix}$, $\kappa_{jy}$ and $\kappa_{ixjy}$ from a data set consisting of an ensemble of events in each of which $x$ and $y$ have been measured.  We can now define the {\it intensive} normalized cumulants that we shall analyze:
\begin{align}
\omega_{i\pi}&\equiv \frac{\kappa_{i\pi}}{\langle N_\pi \rangle }\ ,\label{OmeganPiDefn}\\
\omega_{ip}&\equiv \frac{\kappa_{ip}}{\langle N_p \rangle }\ ,\label{OmeganPDefn}\\
\omega_{i(p-\bar p)}&\equiv \frac{\kappa_{i(p-\bar p)}}{\langle N_{p}+N_{\bar p} \rangle }\ ,\label{OmeganNetPDefn}\\
\omega_{ipj\pi}&\equiv \frac{\kappa_{ipj\pi}}{\langle N_p \rangle^{i/r} \langle N_\pi\rangle^{j/r} }\ ,\label{OmegaipjpiDefn}\\
\omega_{i(p-\bar p)j\pi}&\equiv \frac{\kappa_{i(p-\bar p)j\pi}}{\langle N_{p}+N_{\bar p} \rangle^{i/r} \langle N_\pi\rangle^{j/r} }\ ,\label{OmegaipjNetPDefn}
\end{align}
where $r\equiv i+j$. 

If $N_\pi$, $N_p$ and $N_{\bar p}$ are statistically independent and Gaussian distributed, then the $\omega_2$'s in 
(\ref{OmeganPiDefn}), (\ref{OmeganPDefn}) and (\ref{OmeganNetPDefn}) are nonzero and all the other $\omega$'s vanish.

If $N_\pi$, $N_p$ and $N_{\bar p}$ are statistically independent and Poisson distributed, then all the $\omega_i$'s in 
(\ref{OmeganPiDefn}), (\ref{OmeganPDefn}) and (\ref{OmeganNetPDefn}) with $i\geq 2$ are equal to $1$, and all the mixed cumulants vanish and therefore so do the $\omega$'s in   
(\ref{OmegaipjpiDefn}) and (\ref{OmegaipjNetPDefn}).

In this paper we shall calculate the contributions of critical fluctuations to the normalized cumulants
(\ref{OmeganPiDefn}), (\ref{OmeganPDefn}) and (\ref{OmeganNetPDefn}) for
$i=2$, 3 and 4 and the normalized mixed cumulants 
(\ref{OmegaipjpiDefn}) and (\ref{OmegaipjNetPDefn}) for $i$'s and $j$'s such that $r=2$, 3 and 4.

\subsection{Dependence of $\xi$ on $\mu_B$}
\label{sec:dependence}

We shall close this Introduction (in Section I.C) by illustrating {\em possible} experimental outcomes of measurements of the cumulants defined in Section I.A, 
assuming that the matter produced at the freezeout point of the fireball evolution for some collision energy $\sqrt{s}$ is near the critical point. 
In Section I.C we shall  present only results, while the calculations involved are presented in Section II.   What we shall calculate in Section II is the contribution of critical fluctuations to the observables defined in Section I.A, in terms of the correlation length $\xi$.  In order to give an example of possible experimental outcomes, we need to make an illustrative choice of how 
the correlation length $\xi$ that is achieved in a heavy ion collision depends on $\mu_B$.

To start, let us assume that the critical point occurs at $\mu_B^c=400$ MeV. Let us also assume that because the fireball only spends a finite time in the vicinity of the critical point the correlation length reaches a maximum value of $\xi_{\rm max}=2$~fm in the collisions in which the freeze-out point is closest to the critical point during an energy scan.  We stress that our choices of $\mu_B^c$ and $\xi_{\rm max}$ are arbitrary, made for illustrative purposes only, and are in no way predictions.


How does the correlation length achieved in a heavy ion collision depend on the $\mu_B$ at which the matter produced in the collision freezes out? Close to the critical point, the equilibrium correlation length $\xi_{\rm eq}$ is very long and there is not sufficient time for the actual correlation length $\xi$ achieved in a collision to reach $\xi_{\rm eq}$~\cite{Berdnikov:1999ph}.  Lets suppose that $\xi$ reaches $\xi_{\rm eq}$ for 
$|\mu_B - \mu_B^c| \gtrsim W$, for some $W$, while for 
$|\mu_B - \mu_B^c| \lesssim W$ finite time effects limit $\xi$ such that it peaks at $\xi_{\rm max}$.
In principle, $\xi_{\rm eq}(\mu_B)$ could one day be determined from lattice QCD calculations, but these calculations are challenging at $\mu\neq 0$ because of the notorious fermion sign problem, so this day remains in the future.  At present, all we can do is require
that the static correlation length $\xi_{\rm eq}$ satisfy the constraints imposed by the universality of critical behavior at long wavelengths.  The universal behavior is really only attained in the limit in which $W\rightarrow 0$ and $\xi_{\rm max}\rightarrow\infty$, so our use of it in the present context is illustrative but not quantitative.
As a function of $\mu_B-\mu_B^c$, in the universal regime $\xi_{\rm eq}$ must scale
as $\xi \to f_\pm |\mu_B-\mu_B^c|^{-\nu}$, where $\nu$ is the relevant 
critical exponent\footnote
{ For our illustrative model of the $\xi(\mu_B)$ dependence along the
 freezeout curve we are assuming that where the freezeout curve passes the critical point it is
approximately
 parallel to the transition line (crossover and first-order lines).  The region of the QCD
 phase diagram in the $(\mu_B,T)$ plane near the critical point can
 be mapped onto the Ising model phase diagram, whose reduced
 temperature and magnetic field axes are conventionally denoted by
 $t$ and $h$, respectively.  Upon approaching the Ising critical
 point along the $t$-direction, i.e., along the transition line,
 $\xi_{\rm eq} \sim t^{-\nu}\sim t^{-2/3} $, while along the
 $h$-direction, $\xi_{\rm eq}\sim h^{-\nu/\beta\delta}\sim h^{-2/5}$. As long as
 $h\ll t^{\beta\delta}$  on the freezeout curve, the $t$-like scaling
 dominates and, since $|\mu_B-\mu_B^c|\sim t$, we obtain $\xi_{\rm eq} \sim
 |\mu_B-\mu_B^c|^{-\nu}$. The condition $h\ll t^{\beta\delta}$ is
 violated at points on the freezeout curve that are very close to the critical
 point, $t\approx 0$, where the $h$-like scaling sets in. For
 simplicity we assume that this small-$t$ segment of the freezeout
 curve in the QCD phase diagram lies in a region where the
 equilibrium correlation length $\xi_{\rm eq}$ already exceeds $\xi_{\rm max}=2$ fm, and
 thus $\xi\approx\xi_{\rm max}$ in this segment.  }
and  $f_+$ and $f_-$ are the amplitudes of the singularity on the crossover and first-order side of the transition respectively.
The precise value of the critical exponent is $\nu=(2-\alpha)/3\approx 0.63$, with the numerical value being that for a critical point in the Ising universality class~\cite{Guida:1998bx}. But, in our calculation in Section II we shall be neglecting 
the small anomalous dimensions associated with nonvanishing values of the exponents 
$\eta\approx 0.04$ and $\alpha\approx 0.1$.    So, to be consistent, here too we shall 
simply use $\nu = 2/3$. 
The ratio of the amplitudes $f_+/f_-$ is also a universal quantity. In the Ising universality class, $f_+/f_- \approx 1.9$~\cite{ZinnJustin}.
Since $f_+/f_- > 1$, the correlation length falls off more slowly on the crossover side $\mu<\mu_B^c$.

\begin{figure}[t]
  \centering
  \includegraphics[width=\linewidth]{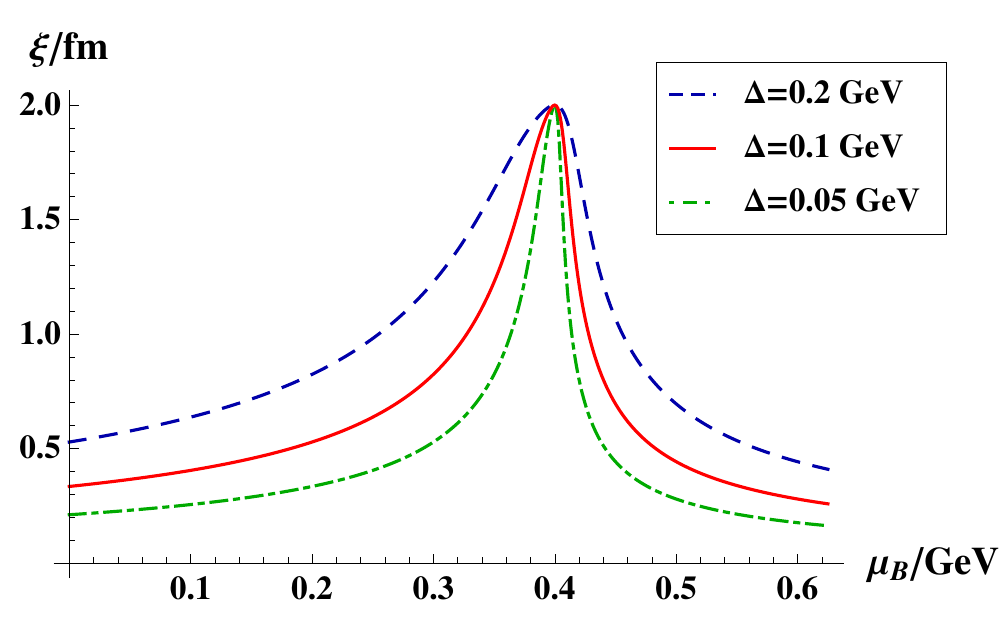}
  \caption{The correlation length $\xi(\mu_B)$ achieved in a heavy ion collision that freezes out with a chemical potential $\mu_B$, according to the ansatz described in the text. We have assumed that the collisions that freeze out closest to the critical point are those that freeze out at  $\mu_B^c=400$~MeV. We have assumed that the finite duration of the collision limits $\xi$ to $\xi<\xi_{\rm max}=2$~fm.  We show $\xi(\mu_B)$ for three choices of the width parameter $\Delta$, defined in the text.  The choices of parameters that have gone into this ansatz are arbitrary, made for illustrative purposes only.  They are not predictions.}
  \label{fig:xi-mub}
\end{figure}

The simplest ansatz for $\xi(\mu_B)$ that we have found that incorporates the physics that we have just described is
\begin{equation}
  \label{eq:xi}
  \xi(\mu_B) = \frac{\xi_{\rm max}}{ \left[ 1 + \frac{(\mu_B-\mu_B^c)^2}{W(\mu_B)^2} \right]^{1/3}}\ ,
\end{equation}
with 
\begin{equation}
W(\mu_B) = W + \delta W \tanh\left(\frac{\mu_B-\mu_B^c}{w}\right)
\end{equation}
where $W$ and $w$ are nonuniversal parameters to be chosen and $\delta W$ is specified by requiring that 
\begin{equation}
\frac{W+\delta W}{W-\delta W} = \left(\frac{f_+}{f_-}\right)^{3/2} = 1.9^{3/2}\ .
\end{equation}
We have constructed (\ref{eq:xi}) such that $\xi$ has the universal behavior of $\xi_{\rm eq}$ when $|\mu_B-\mu_B^c| \gg W(\mu_B)$, but has a peak that is cut off at $\xi=\xi_{\rm max}$ where $\mu_B=\mu_B^c$.  We have chosen the shape of $\xi$ in the vicinity of the peak arbitrarily, for illustrative purposes, not via analysis of the rate of growth of $\xi$ during the finite duration in time of a heavy ion collision.
In Fig.~\ref{fig:xi-mub} we show two instances of our ansatz for $\xi(\mu_B)$.   They differ in their choice of the width of the peak.  We shall define the width $\Delta$ as the distance in $\mu_B$ between the two points at which $\xi(\mu_B)$ crosses 1~fm, i.e. the width in $\mu_B$ within 
which $\xi>1$~fm.  The three curves in the figure have $\Delta$=50, 100 and 200 MeV.  In all three cases we have chosen $w=0.1\Delta$.  (With this choice, $W=0.189 \Delta$ and $\delta W = 0.084 \Delta$.)   There is no reason to expect that $\Delta$ should be small and, indeed, in model calculations it seems to be larger than 100 MeV~\cite{Hatta:2002sj}.  Ultimately $\Delta$ should be determined by lattice calculations; one first attempt to do so indicates 
$\Delta \sim 100$~MeV~\cite{Gavai:2008zr,GuptaPrivate}.

\subsection{Cumulants near the critical point}
\label{sec:fluct-measures}

We shall concentrate our analysis on observables characterizing the fluctuations of pions and protons. Pions are the most abundant species produced in relativistic heavy ion collisions. Protons are important, among other reasons, because their fluctuations are proxy to the fluctuations of the conserved baryon number~\cite{Hatta:2003wn} and because their coupling to the critical mode $\sigma$ is relatively large.

\begin{figure}
  \centering
    \includegraphics*[width=1.03\columnwidth]{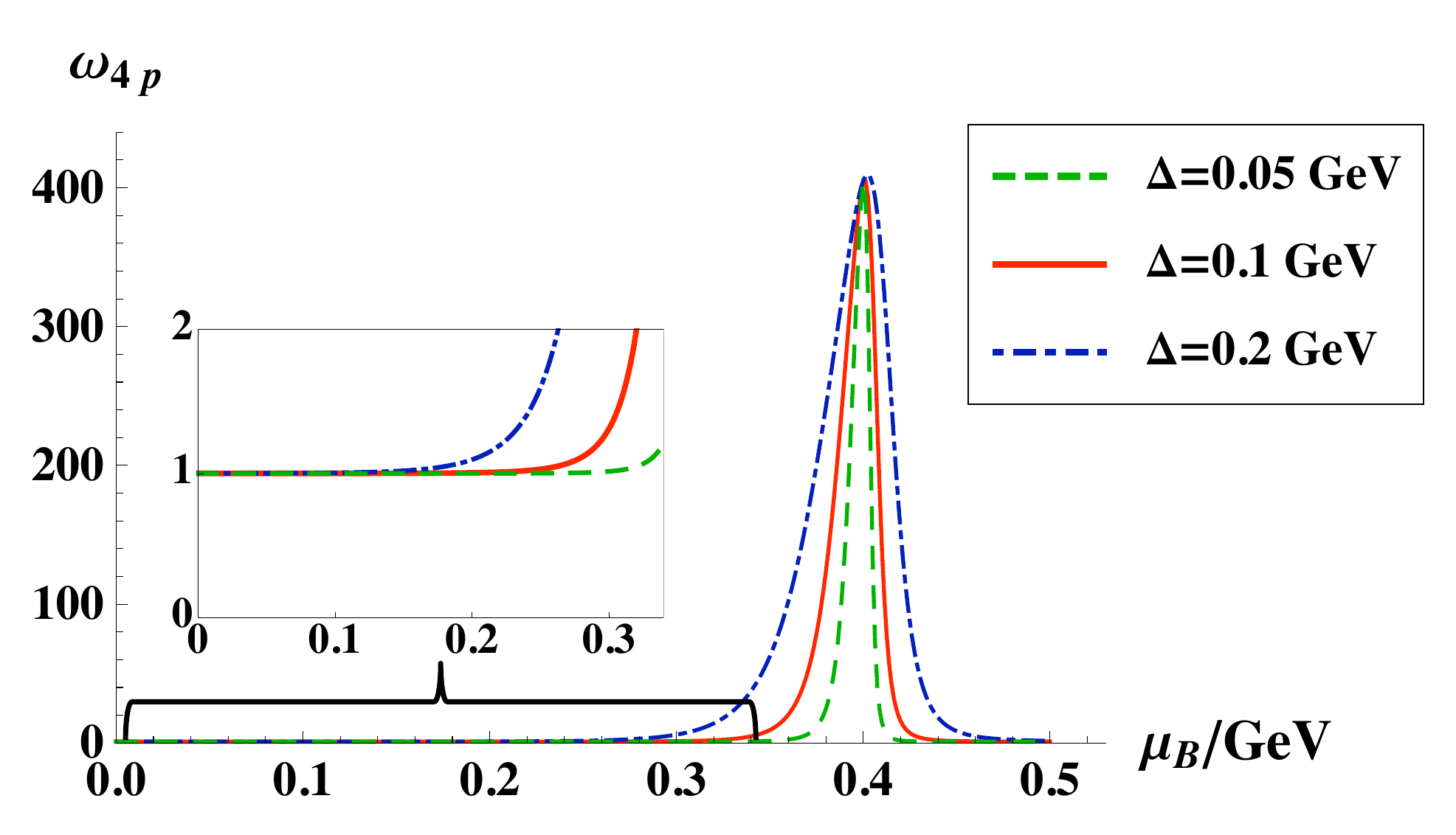}
    \includegraphics*[width=1.01\columnwidth]{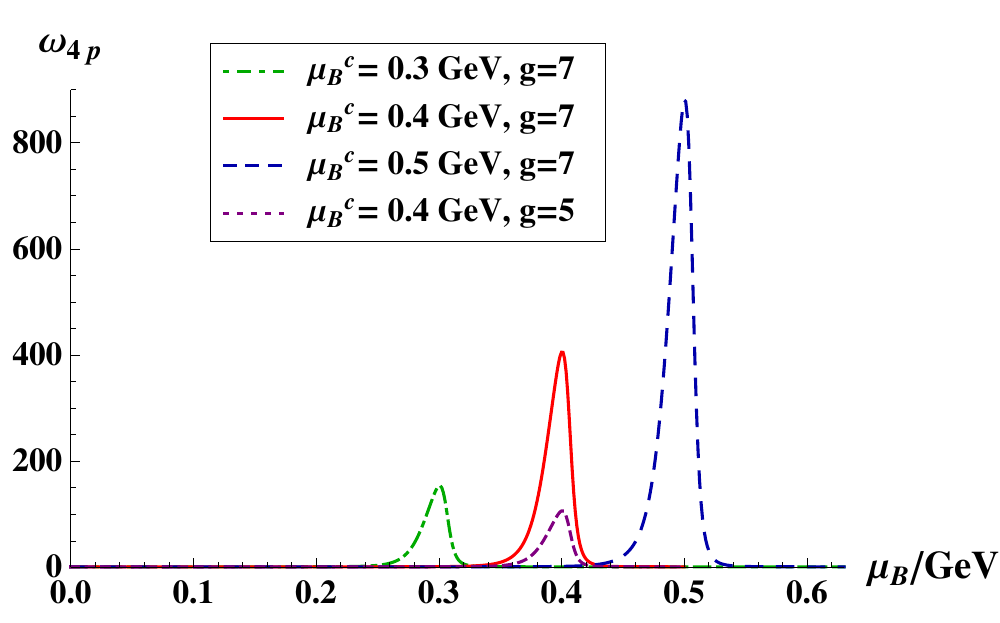}
  \caption{The $\mu_B$-dependence of $\omega_{4p}$, the normalized 4th cumulant of the proton number distribution defined in ({\protect\ref{OmeganPDefn}}), with 
a $\mu_B$-dependent $\xi$ given by ({\protect\ref{eq:xi}}). We only include the Poisson and critical contributions to the cumulant.  In the top panel we choose $\mu_B^c=400$~MeV and 
illustrate how $\omega_{4p}$ is affected if we vary the width $\Delta$ of the peak in $\xi$ from 50 to 100 to 200 MeV, as in
Fig.~{\protect\ref{fig:xi-mub}}.  The inset panel zooms in to show how $\omega_{4p}$ is dominated by the Poisson contribution well below $\mu_B^c$.
In the lower panel, we take $\Delta=100$~MeV and illustrate the effects of changing $\mu_B^c$  and of reducing the sigma-proton coupling $g_p$ from our benchmark $g_p=7$ to $g_p=5$.
}\label{fig:om4pa}
\end{figure}

\begin{figure}
  \centering
    \includegraphics*[width=\columnwidth]{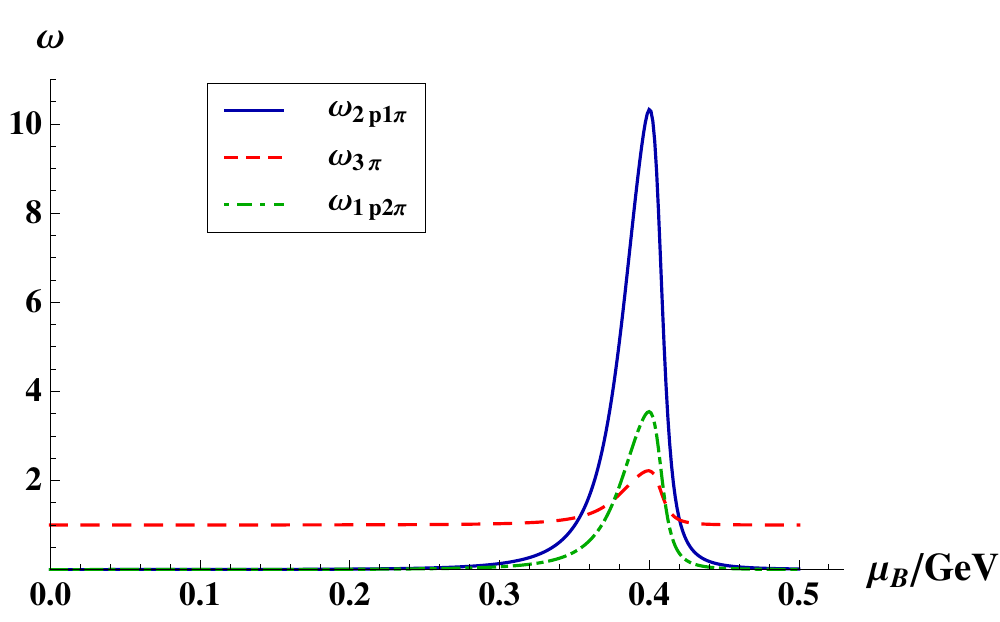}
    \includegraphics*[width=\columnwidth]{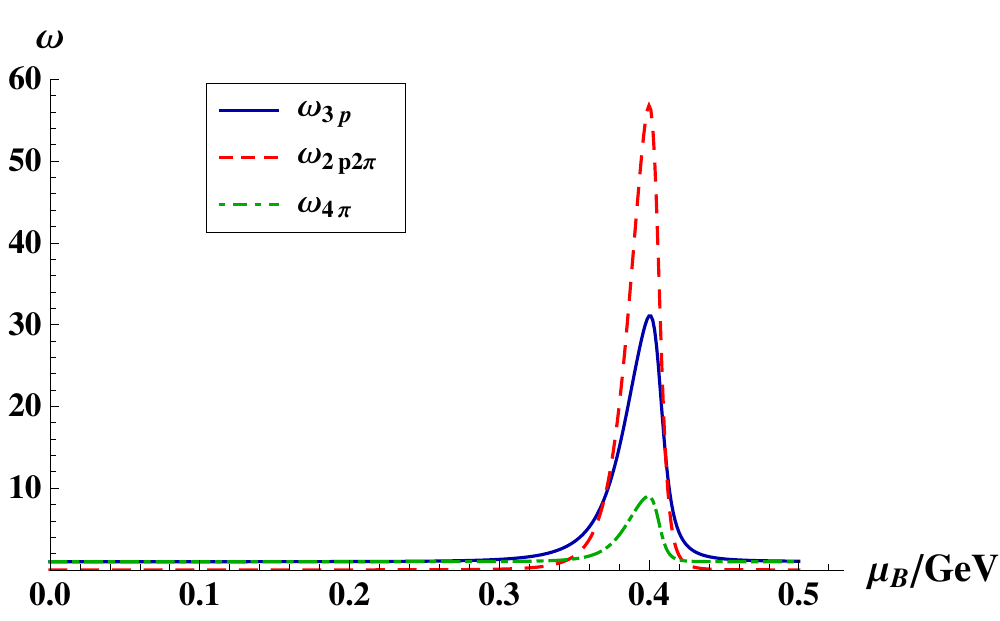}
  \caption{The $\mu_B$-dependence of selected normalized cumulants, defined in ({\protect\ref{OmeganPiDefn}}), ({\protect\ref{OmeganPDefn}}) and ({\protect\ref{OmegaipjpiDefn}}), with a $\mu_B$-dependent $\xi$ given by ({\protect\ref{eq:xi}}) as in Fig.~{\protect\ref{fig:xi-mub}}. We only include the Poisson and critical contributions to the cumulants. We have set all parameters to their benchmark values, described in the text, and we have chosen the width of the peak in $\xi$ to be $\Delta=100$~MeV.  Note the different vertical scales in these figures and in Fig.~\ref{fig:om4pa}; The magnitude of the effect of critical fluctuations on different normalized cumulants differs considerably, as we shall discuss in Sections II and III.  As we shall also discuss in those Sections, ratios of the magnitudes of these different observables depend on (and can be used to constrain) the correlation length $\xi$, the proton number density $n_p$, and four non-universal parameters.  We shall also see in Section III that there are ratios among these observables that are independent of all of these variables, meaning that we can predict them reliably. For example, we shall see that critical fluctuations must yield 
$\omega_{2p2\pi}^2 = (\omega_{4p}-1)(\omega_{4\pi}-1)$ and 
  $\omega_{2p1\pi}^3=(\omega_{3p}-1)^2 (\omega_{3\pi}-1)$ and 
  $\omega_{1p2\pi}^3=(\omega_{3p}-1) (\omega_{3\pi}-1)^2$.
 (The subtractions of 1 are intended to remove the Poisson background; in an analysis of experimental data these subtractions could be done by subtracting the $\omega_{ip}$ or $\omega_{j\pi}$ determined from a sample of mixed events, as this would also subtract various other small background effects.)
}\label{fig:om4pb}
\end{figure}

We have defined the normalized cumulants of the proton and pion distributions in (\ref{OmeganPDefn}) and (\ref{OmeganPiDefn}) and the normalized mixed cumulants in (\ref{OmegaipjpiDefn}).
Fig.~\ref{fig:om4pa} shows how $\omega_{4p}$
might look like, with $\xi(\mu_B)$ given by Eq.~(\ref{eq:xi}). 
We illustrate how $\omega_{4p}$ changes if we vary the location of the critical point $\mu_B^c$ and the width $\Delta$ of the peak in Fig.~\ref{fig:xi-mub}, as well as the sigma-proton 
coupling $g_p$.  As we shall see in Section~\ref{subsection1}, there are four nonuniversal parameters that (for a given $\xi_{\rm max}$) govern the height of the peaks of the normalized cumulants.  These include $g_p$ and the sigma-pion coupling $G$, as well as two parameters $\tilde{\lambda}_3$ and $\tilde{\lambda}_4$ that we shall define in Section~\ref{subsection1}.   
We have used as our benchmark values $G=300$ MeV, $g=7$, $\tilde{\lambda}_3=4$ and $\tilde{\lambda}_4=12$.
As we shall discover in Section II and discuss at length in Section III, the heights of the peaks of different normalized cumulants are affected differently by variations in these four parameters.  
Fig.~\ref{fig:om4pb} shows how six more different normalized cumulants vary with $\mu_B$.  In this figure we keep all parameters set at their benchmark values, deferring a discussion of how these peaks change with parameters to Section III.

In the case of free particles in the classical Boltzmann regime, with no critical fluctuations, the fluctuations of any particle number obey Poisson statistics.  The Poisson contribution to $\omega_{ip}$ and $\omega_{i\pi}$ is 1, and in the figures we have added this Poisson contribution to the contribution from critical flucuations that we calculate in Section II.  There is no Poisson contribution to the mixed cumulants $\omega_{ipj\pi}$.
In reality, in the absence of any critical fluctuations the 1 of Poisson statistics gets few percent contributions from Bose-Einstein statistics, from initial state correlations that are incompletely washed out, and from interactions other than those with the fluctuations that are enhanced near the critical point. 
System size fluctuations are also a potential non-critical contribution to the fluctuation measures. We do not attempt to estimate this effect (see, e.g., Refs.~\cite{Stephanov:1999zu}, \cite{Konchakovski:2008}), and assume that sufficiently tight centrality binning suppresses it.
We are ignoring all of these non-critical corrections to the Poissonian 1 and in the plots shown here we only include the Poisson and critical contributions to the cumulants.   Existing data on $\kappa_{4(p-\bar p)}/\kappa_{2(p-\bar p)}$ at $\sqrt{s}=19.6$, 62.4 and 
200 GeV~\cite{Aggarwal:2010wy} confirm that the non-critical corrections to the Poissonian 1 are indeed small, and confirm that it is possible to measure 4th order cumulants with an error bar that is much smaller than 1.

We can clearly see the peak in all the normalized cumulants near the critical point. In 
many cases, the peak due to critical fluctuations is larger than the Poisson contribution by more than an order of magnitude.\footnote{Although it is a small effect, note that the peaks of any of the cumulants involving protons do not occur exactly at the $\mu_B^c$ at which $\xi(\mu_B)$ from Fig.~\ref{fig:xi-mub} peaks, because the cumulants themselves depend directly on the proton number density and hence on $\mu_B$, as we shall see in Section II.}
The results indicate that the more protons are involved in the observation measure, the easier it is to identify the critical contribution. 
The reader who would like to see an example of a probability distribution that has $\omega_4$ as large as $\omega_{4p}$ gets in Fig.~\ref{fig:om4pa} should consult Appendix A.  
A more comprehensive discussion of the results is given in Sections \ref{section3} and \ref{discussion}, but it is readily apparent that the measurement of these observables in heavy ion collisions at a series of collision energies is very well suited to ruling out (or discovering) the presence of the QCD critical point in the vicinity of the freeze-out points of the collisions in such an energy scan.

\section{Calculating Critical Correlators and Cumulants}
\label{section1}

In this section, we show how to calculate the critical point contribution to the cumulants of the particle multiplicity distribution of pions, protons and net protons. We essentially show how to obtain the normalized cumulants in Figs.~\ref{fig:om4pa} and \ref{fig:om4pb} as the location of the critical point, $\mu_B^c$, changes.  We begin in Section~\ref{subsection1} by calculating the correlators that describe the critical contributions to the fluctuations of the occupation number of 
pions and protons with specified momenta. We use these correlators to calculate the normalized cumulants in Section~\ref{subsection2}.

\subsection{Critical point contribution to correlators}
\label{subsection1}






Fluctuations of observables, such as particle multiplicities, are sensitive to the proximity of the critical point 
if the particles under consideration interact with the critical field $\sigma$ --- the field whose equilibrium correlation length diverges at
the critical point.   In this Section, we shall treat the $\sigma$ correlation length $\xi$ as a parameter, in this way avoiding any consequences of our lack of knowledge of the dynamics of how the long wavelength correlations in the $\sigma$ field grow.   In order to use the results of this section to make the plots in Section I.C, in Section I.B we had to make an ansatz for $\xi(\mu_B)$.  But, the results of this section, expressed in terms of $\xi$, are independent of the uncertainties in that ansatz.

We can describe the fluctuations of the $\sigma$-field by a probability distribution of the form
\begin{equation} \label{eq:Psigma}
P(\sigma) \sim \mathrm{exp} (-\Omega(\sigma)/T),
\end{equation}
where $\Omega$ is the effective action functional for $\sigma$.  It can be expanded in gradients and powers of $\sigma$ as
\begin{equation}\label{eq:Omega}
\Omega(\sigma) = \int d^3x \left[ \frac{1}{2}(\bm\nabla \sigma)^2 + \frac{m_\sigma^2}{2} \sigma^2 +\frac{\lambda_3}{3} \sigma^3 + \frac{\lambda_4}{4} \sigma^4 + ... \right].
\end{equation} 
In this expression the sigma-field screening mass is 
\begin{equation}
m_\sigma \equiv \xi^{-1}
\label{MandXi}
\end{equation}
and, near the critical point, the $\sigma^3$ and $\sigma^4$ interaction couplings are given by
\begin{equation}
\lambda_3 = \widetilde{\lambda}_3 \:T \:(T \:\xi)^{-3/2}, \;\:\:\: \mathrm{and} \:\: \lambda_4 = \widetilde{\lambda}_4  \:(T\: \xi)^{-1},
\label{CriticalLambdas}
\end{equation}
where the dimensionless couplings $\widetilde{\lambda}_3$ and $\widetilde{\lambda}_4$ do not depend on $\xi$, but do depend on the {\em direction} of approach to the critical 
point, as described in Ref.~\cite{Stephanov:2008qz}. 
These couplings (and their dependence on direction) are universal and they have been determined for the Ising universality class~\cite{Tsypin:1994nh}.   Throughout this paper we shall use $\tilde\lambda_3=4$ and $\tilde\lambda_4=12$ as benchmark values, because these are the midpoints of the ranges of values known for these constants~\cite{Tsypin:1994nh,Stephanov:2008qz}.   In fact, both $\tilde\lambda_3$ and $\tilde\lambda_4$ will vary with $\mu_B$, as the location of the freeze-out point moves in the phase diagram, relative to the critical point.   We shall not attempt to parametrize the $\mu_B$-dependence of these parameters, however, because the dominant source of $\mu_B$-dependence in our results is the variation of $\xi$ with $\mu_B$, and our knowledge of $\xi(\mu_B)$ is sufficiently uncertain (as we saw in Section I.B) that this uncertainty would dominate any  increase in precision that would be obtained by modelling the $\mu_B$-dependence of $\tilde\lambda_3$ and $\tilde\lambda_4$.

The correlation functions and fluctuation moments and cumulants of the critical field $\sigma$ itself  can be calculated directly using the probability distribution given in (\ref{eq:Psigma}), but these quantities are not directly observable.   The long wavelength fluctuations in the $\sigma$-field manifest themselves in observable quantities in so far as
they affect the fluctuations of the occupation numbers of particles that couple to the $\sigma$-field. This coupling to the fluctuating field $\sigma$ contributes 
to the moments of particle fluctuations the terms proportional to the corresponding moments of $\sigma$ itself~\cite{Stephanov:2008qz}. 
Both protons and pions couple to the $\sigma$ field. We shall define the strengths of the corresponding couplings $g_p$ and $G$ through the respective terms of the effective Lagrangian (following the notations of \cite{Stephanov:1999zu,Hatta:2003wn}):
\begin{equation}
\mathcal{L}_{\sigma \pi \pi, \sigma pp} = 2 \: G \: \sigma \: \pi^+ \pi^- 
+ g_p\ \sigma\ \bar p\ p.
\end{equation}
where $\pi^\pm$ is the (charged) pion field and $p$ is the Dirac
fermion field of the protons.
The coupling that we denote $g_p$ is often just called $g$.  We shall make the discussion that follows similar for protons and pions by defining a dimensionless measure of the sigma-pion coupling
\begin{equation}
  \label{eq:g-pi}
  g_\pi \equiv G/m_\pi ,
\end{equation}
and using the notation $g$ when we intend an equation to be valid for either pions, with $g\rightarrow g_\pi$, or protons, with $g\rightarrow g_p$.  
Throughout this paper we will use $G = 300$ MeV (see Ref.~\cite{Stephanov:1999zu} for a discussion of how to estimate $G$) and $g_p=7$ 
(see, e.g.,~\cite{Kapusta}) as benchmark values.
It is important to bear in mind that both these parameters and $\tilde\lambda_3$ and $\tilde\lambda_4$ are all uncertain at the  factor of 2 level.  These parameters enter into our calculations of the various normalized multiplicity cumulants, making absolute predictions of these observables in terms of $\xi$ difficult.  The advantage that we have, however, is that we will be able to calculate many different normalized cumulants that depend differently on these parameters.  
In Section~\ref{section3} we shall discuss how to use deliberately chosen  ratios of cumulants to measure and even overconstrain various combinations of these parameters.  And, we shall find five ratios of cumulants that are independent of the values of all of these parameters, allowing us to make parameter-free predictions of these ratios.

The critical contribution to the proton or pion correlators arises from virtual $\sigma$-exchanges which introduce powers of the correlation length $\xi = m_\sigma^{-1}$, where $m_\sigma$ is the $\sigma$-field screening mass. As the correlation length grows in the vicinity of  the critical point, 
the contribution to the particle correlators due to a $\sigma$-exchange dominates over other non-critical contributions. The effect of such an interaction on the two-point particle correlators was studied in Refs.~\cite{Stephanov:1999zu,Stephanov:2001zj} and on higher-point correlators in Ref.~\cite{Stephanov:2008qz}. In this subsection we will only look at the particle correlators and in the subsequent sections we will show how to calculate cumulants of particle multiplicity distributions from the correlators. 

The contribution of critical fluctuations to the 2-, 3- and 4- particle correlators due to $\sigma$-exchanges can be calculated using the diagrammatic method developed in Ref.~\cite{Stephanov:2001zj} (see also Refs.~\cite{Stephanov:1999zu} and \cite{Stephanov:2008qz}).  We shall write the correlators using a notation that applies to either protons or pions. They describe the correlation between the $\delta n_\mathbf{\p}$'s at different momenta, where $\delta n_\mathbf{\p} \equiv n_\mathbf{\p}  - \langle n_\mathbf{\p}\rangle$ is the difference between the occupation number of  the $\mathbf{\p}$'th pion or proton mode in momentum space in a particular event and its mean value.  The correlators are given by
\begin{eqnarray} \label{eq:2corr}
 \langle \delta n_\mathbf{\p_1} \delta n_\mathbf{\p_2} \rangle_{\sigma} &=& \frac{d^2}{m_{\sigma}^2 V}\frac{g^2}{T} \frac{v^2_\mathbf{\p_1}}{\omg_\mathbf{\p_1}}  \frac{v^2_\mathbf{\p_2}}{\omg_\mathbf{\p_2}} \nonumber\\
&=&  \frac{d^2}{VT}\, g^2\xi^2\frac{v^2_\mathbf{\p_1}}{\omg_\mathbf{\p_1}}  \frac{v^2_\mathbf{\p_2}}{\omg_\mathbf{\p_2}} \ ,
 \end{eqnarray}
\begin{eqnarray}  \label{eq:3corr}
 \langle \delta n_\mathbf{\p_1} \delta n_\mathbf{\p_2} \delta n_\mathbf{\p_3}  \rangle_{\sigma} &=&  \frac{2 d^3 \lambda_3}{V^2 T} \left(\frac{g}{m^2_\sigma}\right)^3 \frac{v^2_\mathbf{\p_1}}{\omg_\mathbf{\p_1}}  \frac{v^2_\mathbf{\p_2}}{\omg_\mathbf{\p_2}}  \frac{v^2_\mathbf{\p_3}}{\omg_\mathbf{\p_3}} \nonumber\\
&=& 
\frac{2d^3\tilde\lambda_3}{V^2 T^{3/2}} \, g^3 \xi^{9/2} \frac{v^2_\mathbf{\p_1}}{\omg_\mathbf{\p_1}}  \frac{v^2_\mathbf{\p_2}}{\omg_\mathbf{\p_2}}  \frac{v^2_\mathbf{\p_3}}{\omg_\mathbf{\p_3}} ,\quad
\end{eqnarray}
\begin{align} \label{eq:4corr}
\langle \langle &\delta n_\mathbf{\p_1} \delta n_\mathbf{\p_2} \delta n_\mathbf{\p_3}  \delta  n_\mathbf{\p_4} \rangle \rangle_\sigma\notag\\  
&=  \frac{6d^4}{V^3 T} \left(2 \left(\frac{\lambda_3}{m_\sigma} \right)^2 - \lambda_4 \right) 
\left(\frac{g}{m^2_\sigma}\right)^4 \frac{v^2_\mathbf{\p_1}}{\omg_\mathbf{\p_1}}  \frac{v^2_\mathbf{\p_2}}{\omg_\mathbf{\p_2}}  \frac{v^2_\mathbf{\p_3}}{\omg_\mathbf{\p_3}}  \frac{v^2_\mathbf{\p_4}}{\omg_\mathbf{\p_4}}\notag \\ 
&=
 \frac{6d^4}{V^3 T^2} \left(2 \tilde\lambda_3^2 - \tilde\lambda_4 \right)g^4 \xi^7\,
 \frac{v^2_\mathbf{\p_1}}{\omg_\mathbf{\p_1}}  \frac{v^2_\mathbf{\p_2}}{\omg_\mathbf{\p_2}}  \frac{v^2_\mathbf{\p_3}}{\omg_\mathbf{\p_3}}  \frac{v^2_\mathbf{\p_4}}{\omg_\mathbf{\p_4}}\ ,
\end{align}
where we have used (\ref{MandXi}) and (\ref{CriticalLambdas}) and where we must now explain many aspects of our notation.  The subscript $\sigma$ indicates that we have only calculated the contribution of the critical fluctuations to the correlators.
The double brackets around the quartic correlator indicate that what is evaluated is the cumulant, as in (\ref{eq:4x}).
The equations~(\ref{eq:2corr}),~(\ref{eq:3corr}),~(\ref{eq:4corr}) apply to both protons (with $g=g_p$) and pions (with $g=g_\pi=G/m_\pi$). 
The degeneracy factor $d$ is 2 for both protons and pions.  (For protons, $d=d_p=2$ counts the number of spin states.  For pions, $d=d_\pi=2$ counts the number of charge states --- $\pi^+$ and $\pi^-$. These degeneracy factors appear because the coupling to the $\sigma$-field is both spin and charge ``blind''.)
The variance of the fluctuating occupation number distribution is denoted by $v^2_\mathbf{\p}$ and is given by
\begin{equation}
v^2_\mathbf{\p}=\langle n_\mathbf{\p} \rangle \left(1 \pm \langle n_\mathbf{\p}\rangle \right),\label{eq:v-2-p}
\end{equation}
where, as usual,
\begin{equation}
\langle n_\mathbf{\p} \rangle = \frac{1}{ \exp \left[  (   \gamma_\mathbf{\p} m - \mu )
 /T \right]  \mp 1
}
\end{equation}
with $m=m_\pi$, $\mu=0$ and the upper sign for pions and $m=m_p$, $\mu=\mu_B$ and the lower sign for protons.
And, finally,
\begin{equation}
\omg_\mathbf{\p} \equiv\frac{ \sqrt{\mathbf{\p}^2+m^2}}{m}
\end{equation}
 is the relativistic gamma-factor of the particle with mass $m$ with a given momentum $\mathbf{\p}$.\footnote{A note on subscript/superscript notation: we denote momentum subscripts with a bold letter $\mathbf{\p}$. Subscripts/superscripts denoting particle type, e.g. $p$ for protons, will be in normal typeface. 
} 
We see  from Eqs.~(\ref{eq:2corr})-(\ref{eq:4corr}) that these correlators, and hence the cumulants that we will obtain from them, are proportional to powers of the correlation length $\xi$ and so peak  at the critical point.

Now let us turn to mixed pion-proton correlators. The 2 pion - 2 proton correlator
is given by
\begin{align} \label{final2p2pi}
\langle \langle& \delta n_\mathbf{\p_1}^{\pi} \delta n_\mathbf{\p_2}^{\pi} \delta n_\mathbf{\p_3}^{p}  \delta n_\mathbf{\p_4}^{p} \rangle \rangle_\sigma \notag\\
&=\frac{6d_\pi^2 d_p^2}{V^3 T} \left(2\left(\frac{\lambda_3}{m_\sigma}\right)^2 -\lambda_4\right)  
\left(\frac{g_\pi\:g_p}{m_\sigma^4}\right)^2 \frac{v_\mathbf{\p_1}^{\pi\:2}}{\omg_\mathbf{\p_1}^\pi} \frac{v_\mathbf{\p_2}^{\pi\:2}}{\omg_\mathbf{\p_2}^\pi} \frac{v_\mathbf{\p_3}^{p\:2}}{\omg_\mathbf{\p_3}^p} \frac{v_\mathbf{\p_4}^{p\:2}}{\omg_\mathbf{\p_4}^p}\notag\\
&=
\frac{6d_\pi^2 d_p^2}{V^3 T^2} \left(2\tilde\lambda_3^2-\tilde\lambda_4\right)g_\pi^2g_p^2 \xi^7 \, 
\frac{v_\mathbf{\p_1}^{\pi\:2}}{\omg_\mathbf{\p_1}^\pi} \frac{v_\mathbf{\p_2}^{\pi\:2}}{\omg_\mathbf{\p_2}^\pi} \frac{v_\mathbf{\p_3}^{p\:2}}{\omg_\mathbf{\p_3}^p} \frac{v_\mathbf{\p_4}^{p\:2}}{\omg_\mathbf{\p_4}^p}.
\end{align}
The prescription for obtaining other mixed correlators from the
correlators (\ref{eq:2corr} - \ref{eq:4corr}) should be clear:
each particle brings its own corresponding factor $d\, g\, v_\mathbf{\p}^2/\omg_\mathbf{\p}$ to the expression in, e.g., Eq.~(\ref{eq:4corr}).  In this way, the 1 pion - 3 proton and 3 pion - 1 proton mixed correlators can be obtained from Eq.~(\ref{eq:4corr}), the 1 pion - 2 proton and 2 pion - 1 proton mixed correlators can be obtained from Eq.~(\ref{eq:3corr}), and the 1 pion - 1 proton can be obtained from Eq.~(\ref{eq:2corr}).


Another useful fluctuating quantity to consider is the {\em net}
proton number correlators (the net proton number is defined as the number of protons minus the number of anti-protons: $N_{p-\bar p} = N_p-N_{\bar p}$). In order to obtain the corresponding correlators one can begin with the similar correlators for the protons and replace $v_\mathbf{\p}^{p\:2}$ with  $(v_\mathbf{\p}^{p\:2}-v_\mathbf{\p}^{\bar p\:2})$, where 
$v_\mathbf{\p}^{\bar p\:2}$ is the occupation number variance for
anti-protons. (See, e.g., Ref.~\cite{Hatta:2003wn}).

In the next section we will use these correlators to evaluate cumulants of particle multiplicity distributions for pions, protons and net protons and see how they can be used to locate the critical point.

\subsection{Energy dependence of pion, proton, net proton, and mixed pion/proton multiplicity cumulants}
\label{subsection2}








In this section we will concentrate on cumulants of the particle multiplicity distributions and how they vary as we change the location of the critical point and change the value of parameters. Another application of the correlators given in the previous section is the calculation of the critical point effect on higher moments of the fluctuation of mean transverse momentum $p_T$. We find that the critical contribution to $p_T$ fluctuations is rather small (e.g., smaller than the enhancement due to Bose statistics) and thus not as useful in the search of the critical point. Details can be found in Appendix \ref{appendixpT}.

Now let us focus on how one can obtain higher cumulants of the particle multiplicity distributions using the correlators found in the previous section. As an example, let us evaluate the critical contribution to the normalized fourth cumulant of the proton multiplicity distribution, $\omega_{4p}$ defined in (\ref{OmeganPDefn}).
The total multiplicity $N_p$ is just the sum of all occupation numbers $n_\mathbf{\p}$, thus (see  ref. \cite{Stephanov:2008qz})
\begin{eqnarray} \label{eq:kappa2pi}
\kappa_{4p, \sigma}&=&\langle \langle (\delta N_p)^4 \rangle\rangle_\sigma\\
 &=& 
V^4\int_\mathbf{\p_1} \int_\mathbf{\p_2} 
\int_\mathbf{\p_3} \int_\mathbf{\p_4} 
\langle\langle \delta n_\mathbf{\p_1}^p \delta n_\mathbf{\p_2}^p  \delta n_\mathbf{\p_3}^p \delta n_\mathbf{\p_4}^p  \rangle \rangle_{\sigma}\ ,\qquad\nonumber
\end{eqnarray}
where
\begin{equation}
 \int_\mathbf{\p} \equiv  \int \frac{d^3 \mathbf{\p}}{(2\pi)^3}.
 \end{equation}
As we discussed in Section I.A, see (\ref{OmeganPDefn}), we normalize the cumulant by dividing by
the total proton multiplicity $N_p$.  To simplify notation below, it is convenient to introduce the proton and pion number densities
\begin{eqnarray}
n_p \equiv \frac{\langle N_p\rangle}{V} &=& d_p \int_\mathbf{\p} \langle n_\mathbf{\p}^p \rangle\nonumber\\
&=&\frac{1}{\pi^2}\int_{m_p}^\infty \frac{ dE\, E \sqrt{E^2-m_p^2}}{e^{(E-\mu_B)/T}+1}
 \label{npDefn}\\
n_\pi \equiv \frac{\langle N_\pi\rangle}{V} &=& d_\pi \int_\mathbf{\p} \langle n_\mathbf{\p}^\pi \rangle\nonumber\\
&=&\frac{1}{\pi^2}\int_{m_\pi}^\infty \frac{ dE\, E \sqrt{E^2-m_\pi^2}}{e^{E/T}-1}
\ .
\end{eqnarray}
The result we find for the normalized cumulant can then be written as
%
\begin{equation}
\omega_{4 p, \:\sigma} =\frac{6\,(2\tilde\lambda_3^2-\tilde\lambda_4)}{T^2 n_p}  \,\xi^7 \left(d_p\, g_p \int_\mathbf{\p} \frac{v^{p\:2}_\mathbf{\p}}{\omg^p_\mathbf{\p}} \right)^4 \ .
\end{equation}
We can see from expressions (\ref{eq:2corr}) - (\ref{eq:4corr}) that higher cumulants are proportional to higher powers of $\xi$ and thus increase by a larger factor near the critical point where $\xi$ becomes large. For example, the third and fourth cumulants are proportional to $\xi^{9/2}$ and $\xi^7$, respectively.  If the correlation length $\xi$ increases from $\sim 0.5$~fm to $\xi_{\rm max}=2$~fm as in Section I.B, these cumulants are substantially enhanced --- as we have seen in the plots in Section I.C.

\begin{figure}
  \centering
  \includegraphics*[width=\columnwidth]{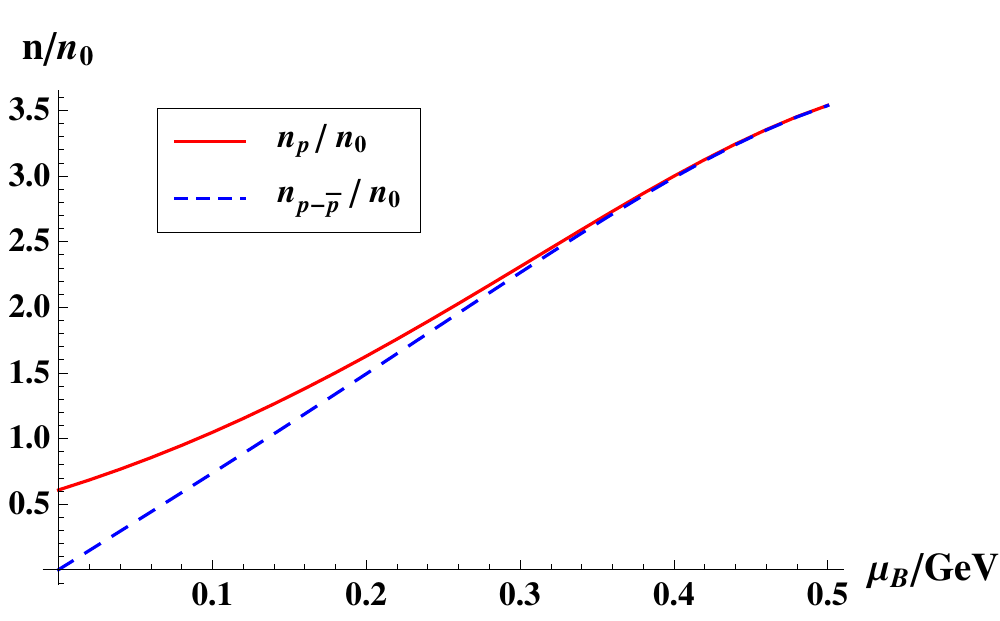}
  \caption{Proton number density $n_p$ and net proton number density $n_{p-\bar p}\equiv n_p - n_{\bar p}$ at chemical freezeout as functions of $\mu_B$.   Both depend on $T$ as well as $\mu_B$; we have taken $T(\mu_B)$ as in (\ref{Tvsmub}). We have normalized both $n_p$ and $n_{p-\bar p}$ using the constant $n_0$ of (\ref{n0Defn}) introduced in (\ref{eq:om}) and (\ref{eq:prefact}). }
  \label{fig:np-mub}
\end{figure}

\begin{figure}
  \centering
    \includegraphics*[width=\columnwidth]{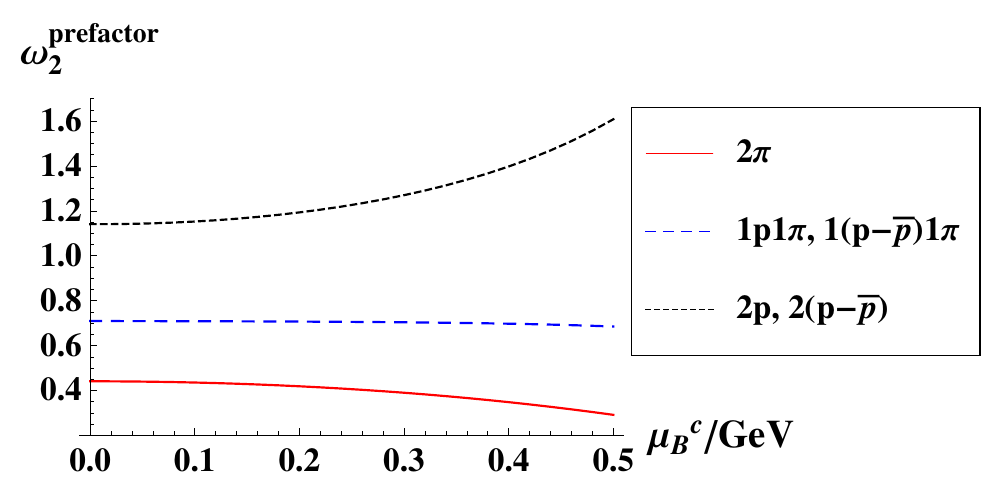}
 \includegraphics*[width=\columnwidth]{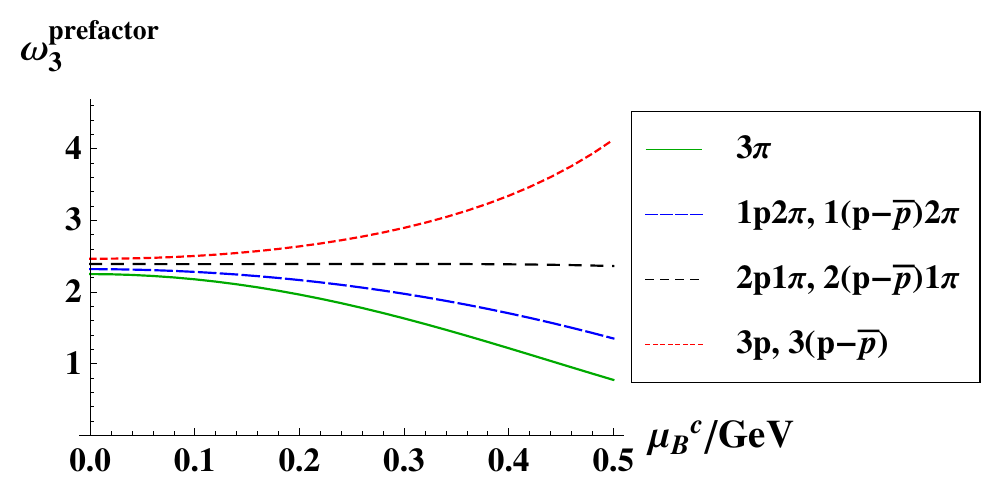}
\includegraphics*[width=\columnwidth]{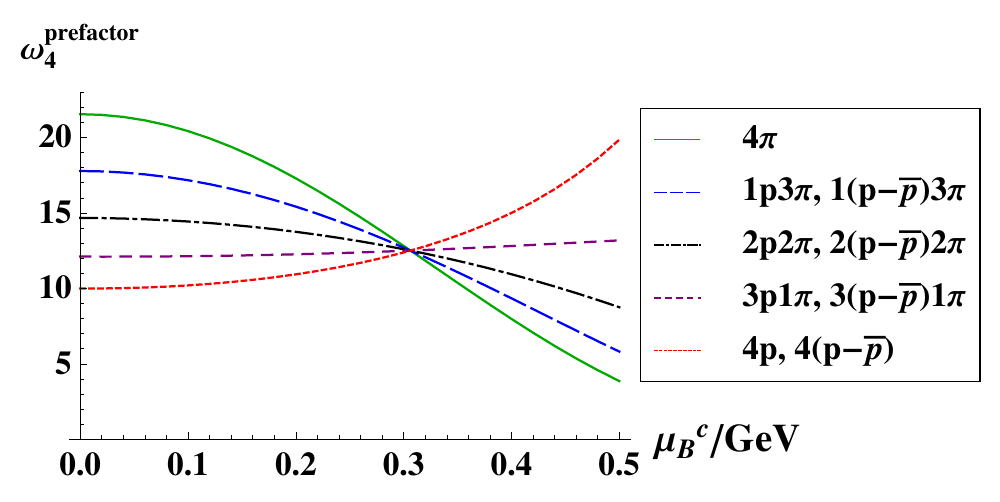}
  \caption{The $\mu_B$-dependence of $\omega_{ipj\pi}^{\rm{prefactor}}$ and $\omega_{i(p-\overline{p})j\pi}^{\rm{prefactor}}$, defined in (\ref{eq:om}), (\ref{eq:prefact}) and (\ref{eq:om2}). The three panels are for the normalized cumulants with $\ipj\equiv i+j=2$, 3 and 4, respectively. The curves can be used to determine how the height of the peak in the critical contribution to the normalized cumulants changes as we vary $\mu_B^c$, the $\mu_B$ at which $\xi=\xi_{\rm max}$ and at which (to a very good approximation) the normalized cumulant has its peak. 
The height of the peak in $\omega_{ipj\pi}$  [or $\omega_{i(p-\bar p)j\pi}$] 
 is proportional to  $(n_p/n_0)^{i-i/r}$ [or $(n_{p-\bar p}/n_0)^{i-i/r}$] multiplied by the prefactor plotted in this Figure. We have taken $T(\mu_B)$ as in (\ref{Tvsmub}) and have used the benchmark parameters 
 $G=300$ MeV, $g_p=7$, $\tilde{\lambda}_3=4$ and $\tilde{\lambda}_4=12$.
    }\label{fig:om2}
\end{figure}

With an explicit expression for $\omega_{4p,\sigma}$ in hand, we can now write our general result for $\omega_{ipj\pi,\sigma}$ in (\ref{OmegaipjpiDefn}). We can also include $\omega_{ip,\sigma}$ and $\omega_{j\pi,\sigma}$ defined as in (\ref{OmeganPDefn})   and (\ref{OmeganPiDefn}) in the notation via setting $j=0$ or $i=0$ in $\omega_{ipj\pi,\sigma}$. We obtain
\begin{eqnarray} \label{eq:om}
\omega_{ipj\pi} &=&
 \delta_{i,0}+\delta_{j,0}+ \frac{\tilde{\lambda}'_{\ipj}\,(r-1)! }{T^{\pipj /2}}\, \frac{\alpha_p^i }{ n_p^{i/\ipj} } \,
\frac{\alpha_\pi^j}{n_\pi^{j/\ipj}}\, \xi^{\frac{5}{2}\pipj  - 3}\\
&=& \delta_{i,0}+\delta_{j,0}+\omega_{ipj\pi} ^{\mathrm{prefactor}} \left(\frac{n_p}{n_0} \right)^{i-\frac{i}{\ipj}}
\left(\frac{\xi}{\xi_{\rm max}}\right)^{\frac{5}{2}\pipj  - 3},\nonumber
\end{eqnarray}
where we have defined
\begin{equation}\label{eq:prefact}
\omega_{ipj\pi} ^{\mathrm{prefactor}} \equiv \frac{\tilde{\lambda}'_{\ipj} \:(r-1)!\:\xi_{\rm max} ^{\frac{5}{2}\pipj  - 3}  }{T^{\pipj /2}} \frac{\alpha_p^i }{ n_p^{i/\ipj} } \,
\frac{\alpha_\pi^j}{n_\pi^{j/\ipj}}\,
 \left(\frac{n_0}{n_p} \right)^{i-\frac{i}{\ipj}}
\end{equation}
and 
\begin{eqnarray}
\alpha_\pi &\equiv& d_\pi\: g_\pi \int_\mathbf{\p} \frac{v_\mathbf{\p}^{\pi\:2}}{\omg_\mathbf{\p}^\pi},  \:\:\:\: \alpha_p \equiv d_p\: g_p  \int_\mathbf{\p} \frac{v_\mathbf{\p}^{p\:2}}{\omg_\mathbf{\p}^p} , \label{eq:alpha} \\
\tilde{\lambda}'_2 &\equiv& 1,\:\:\:  \tilde{\lambda}'_3 \equiv \tilde{\lambda}_3 \:\:\: \mathrm{and}  \:\:\:   \tilde{\lambda}'_4 \equiv 2 \tilde{\lambda}_3^2-\tilde{\lambda}_4.\label{eq:lambda_r}
\end{eqnarray}
In the second line of (\ref{eq:om}) we have factored out the two main sources of $\mu_B$ dependence: the correlation length $\xi$ depends on $\mu_B$ as we have discussed at length in Section IB and, if the normalized cumulant involves the proton multiplicity it depends on $n_p$, which increases rapidly with increasing $\mu_B$ as shown in Fig.~\ref{fig:np-mub}.  We have denoted all of the remaining factors in our result for the contribution of critical fluctuations to the normalized cumulant by 
$\omega_{ipj\pi}^{\rm prefactor}$, which depends only weakly on $\mu_B$ as we illustrate in Fig.~\ref{fig:om2}.   The number density 
$n_0$ is an arbitrary constant --- note that it cancels when (\ref{eq:prefact}) 
is substituted into (\ref{eq:om}) --- introduced in order to make 
$\omega_{ipj\pi}^{\rm prefactor}$ dimensionless.  We shall choose
\begin{equation}
n_0\equiv\frac{1}{(5\, {\rm fm})^3}=6.116\times 10^{-5}~{\rm GeV}^3\ .
\label{n0Defn}
\end{equation}
  With this choice, $\langle n_p\rangle/n_0$ is of order 1 at the $\mu_B$ of interest to us --- see Fig.~\ref{fig:np-mub} --- and none of the different
$\omega_{ipj\pi}^{\rm prefactor}$s are orders of magnitude smaller or larger than 1, as illustrated in Fig.~\ref{fig:om2}.

Let us now walk through the physics behind the  different pieces of the expression (\ref{eq:om}). 
The Kronecker deltas describe Poisson fluctuations, which are of course $\xi$-independent.  As we described in Section IA, they contribute 1 to the $\omega_{ip}$'s and the $\omega_{j\pi}$'s and they make no contribution to the mixed cumulants in which $i$ and $j$ are both nonzero.
More realistically, the 1 of Poisson statistics gets few percent contributions from Bose-Einstein statistics (which are calculable), from initial state correlations that are incompletely washed out, and from interactions other than those with the critical $\sigma$-mode. We are ignoring all of these noncritical corrections to the 1. In principle, with sufficiently precise data their magnitude could be measured far away from the critical point and this background could then be subtracted. If this background were significant, one could also try to study and calculate these corrections theoretically.  Present data on $\kappa_{4(p-\bar p)}/\kappa_{2(p-\bar p)}$ at $\sqrt{s}=19.6$, 62.4 and 200 GeV indicate that the corrections to the Poissonian 1 are quite small, but this should be investigated also for other cumulants.

The second, $\xi$-dependent, term in (\ref{eq:om}) is the contribution to $\omega_{ipj\pi}$ made by the critical fluctuations. It grows proportional to $\xi^{(5\pipj-6)/2}$ near the critical point. 
We see evidence of this in the heights of the peaks in different $\omega$'s in Fig.~\ref{fig:om4pb}, but it is also clear from this Figure that the $r$-dependent difference in the power of $\xi$ is not the only important source of $\mu$-dependence.  Indeed, we see in (\ref{eq:prefact}) that $\omega_{ipj\pi}^{\rm prefact}$ is proportional to $n_p^{-i/r}$ and to $\alpha_p^i$ and, it turns out, $\alpha_p/n_p$ is close to constant.  This means that the dominant $\mu_B$-dependence of the critical contribution to $\omega_{ipj\pi}$ at a given $\xi$ is 
$n_p^{i-i/r}$, 
which we have factored out in (\ref{eq:om}) making the $\mu_B$-dependence in $\omega_{ipj\pi}^{\rm prefact}$ rather mild.
We can see the $n_p^{i-i/r}=n_p^3$ dependence of the height of the peak in $\omega_{4p}$ in the lower panel of Fig.~\ref{fig:om4pa}: in this figure $\xi_{\rm max}$ is the same for all the curves so the $\mu_B$-dependence of the height of the peaks in $\omega_{4p}$ comes from its $n_p$-dependence.

For $i=0$, meaning for a cumulant involving pions only, there is no large $n_p$-dependence in $\omega_{j\pi}$ and the height of the peak in a figure like Fig.~\ref{fig:om4pb} is proportional to $\omega_{j\pi}^{\rm prefactor}$, and the dominant $\mu_B$-dependence of $\omega_{j\pi}$ itself comes from its $\xi^{j-1}$ dependence.
For observables involving protons ($i>0$),
the dominant contribution to the $\mu_B^c$ dependence of the height of the peak in $\omega$ comes from the factor  $n_p^{i-i/\ipj}$, and the slowly varying prefactor in Fig.~\ref{fig:om2} adds relatively little to that strong dependence.

 %
%


In plotting the curves in Fig.~\ref{fig:np-mub} and Fig.~\ref{fig:om2}, we have allowed for the fact that the chemical freeze-out temperature $T$ decreases somewhat with increasing $\mu_B$.  We have described this dependence using an empirical parametrization of heavy ion collision data from Ref.~\cite{Cleymans:2005xv}:
\begin{equation}
T(\mu_B)=a-b \mu^2_B-c \mu_B^4,
\label{Tvsmub}
\end{equation}
with $a=0.166$ GeV, $b=0.139$ GeV$^{-1}$ and $c=0.053$ GeV$^{-3}$. 
Almost all of the $\mu_B$-dependence of the $\omega_{ipj\pi}^{\rm prefactor}$s plotted in Fig.~\ref{fig:om2} actually comes from the $\mu_B$-dependence of the chemical freeze-out temperature $T$. In plotting Fig.~\ref{fig:om2}, we have used our benchmark values of the four nonuniversal parameters that determine the $\omega_{ipj\pi}$ for a given $\xi$, namely
$g_\pi=G/ m_\pi=2.1$, $g_p=7$,  $\widetilde{\lambda}_3=4$ and $\widetilde{\lambda}_4=12$.

Finally, completing our discussion of the proton-pion cumulants and Fig.~\ref{fig:om2}, we note that in the lower panel in this figure there is a point where all five $\omega_{ipj\pi}^{\rm prefactor}$s with $r=4$ cross.
This occurs if at some value of $\mu_B$ it so happens that $\alpha_\pi(n_0/n_\pi)^{1/r}$ and  
$\alpha_p n_0/n_p$ coincide.

We now turn to the net proton multiplicity distribution, where by net protons we mean $N_{p-\bar p}\equiv N_p-N_{\bar p}$.  The calculation of the normalized cumulants involving the net proton multiplicity and the pion multiplicity, namely (\ref{OmegaipjNetPDefn}), is analogous to the calculation we have described above. As we discussed in Section IIA, the only change in the correlator from which the cumulants are obtained is the replacement of 
$v_{\mathbf{\p}}^{p\: 2}$ with $v_{\mathbf{\p}}^{p\: 2} - v_{\mathbf{\p}}^{\bar p\: 2}$.
We find
%
\begin{align} \label{eq:om2}
\omega_{i(p-\overline{p})j\pi} = \:& \delta_{i,0}+\delta_{j,0} + \omega_{i(p-\overline{p})j\pi} ^{\mathrm{prefactor}} \left(\frac{n_{p-\overline{p}}}{n_0} \right)^{i-\frac{i}{\ipj}}  \notag \\
&\times \left(\frac{n_{p-\overline{p}}}{n_p + n_{\overline{p}}} \right)^{\frac{i}{\ipj}}
\left(\frac{\xi}{\xi_{\rm max}}\right)^{\frac{5}{2}\pipj  - 3}
,
\end{align}
where $n_{p-\overline{p}} = n_p - n_{\overline{p}}$ is the net proton number density. In comparison with Eq.~(\ref{eq:om}), we have pulled out another factor, 
$\left(\frac{n_{p-\overline{p}}}{n_p + n_{\overline{p}}} \right)^{\frac{i}{\ipj}}$, 
which describes the vanishing of the critical contribution to net proton cumulants at $\mu_B=0$, see Fig.~\ref{fig:np-mub}.   It then turns out that the 
prefactor $\omega_{i(p-\overline{p})j\pi} ^{\mathrm{prefactor}}$ (defined as  in (\ref{eq:prefact}),   but multiplied by 
$\left(\frac {n_p + n_{\overline{p}}}{n_{p-\overline{p}}} \right)^{\frac{i}{\ipj}}$ and with $v_{\mathbf{\p}}^{p\: 2}$ replaced by $v_{\mathbf{\p}}^{p\: 2} - v_{\mathbf{\p}}^{\bar p\: 2}$)
differs from $\omega_{ipj\pi} ^{\mathrm{prefactor}} $ by less than half of one percent, which is less than the thickness of the curves in Fig.~\ref{fig:om2}. Hence, these curves also depict
$\omega_{i(p-\overline{p})j\pi} ^{\mathrm{prefactor}}$.

In order to evaluate  either (\ref{eq:om}) [or (\ref{eq:om2})] and compare to data, we need
the proton number density $n_p$ [net proton number density $n_{p-\overline{p}}$]
at each collision energy $\sqrt{s}$. These can be extracted from data via the
conventional
statistical model fits done at each $\sqrt{s}$ that give $\mu_B$ and $T$ at chemical freeze-out at each $\sqrt{s}$.  The value of $n_p$  at chemical freeze-out is specified in terms of $\mu_B$ and $T$ by (\ref{npDefn}) and the value of $n_{\bar p}$ is given by the same expression with $\mu_B$ replaced by $-\mu_B$, so these number densities can also be obtained from data.
So, at each collision energy, one should take the $\mu_B$ and $T$ from the statistical model fit, evaluate $n_p$ and $n_{p-\overline{p}}$, and then plug these into (\ref{eq:om}) and (\ref{eq:om2}) and see what conclusions can be drawn about $\xi$ and the constants $g_p$, $g_\pi$, $\tilde\lambda_3$ and $\tilde\lambda_4$ using data on as many of the normalized cumulants $\omega_{ipj\pi}$ and $\omega_{i(p-\overline{p})j\pi}$ as possible.    We shall provide tuned strategies for this analysis in Section \ref{section3}.  We close this Section with two straightforward observations.

First, the proton/pion normalized cumulant $\omega_{ipj\pi}$ is more sensitive to critical fluctuations than the net-proton/pion normalized cumulant $\omega_{i(p-\bar p)j\pi}$, for any $i\neq 0$ and for any $j$.   As an example let us consider $\omega_{4p}$ and $\omega_{4(p-\overline{p})}$. We can estimate $\mu_B(\sqrt{s})$ using the parametrization of statistical model fits to data in Ref.~\cite{Cleymans:2005xv}:
\begin{equation}
\mu_B(\sqrt{s})=\frac{d}{1+e\sqrt{s}},
\end{equation}
with $d=1.308$ GeV and $e=0.273$ GeV$^{-1}$.  The proton number density $n_p(\mu_B)$ is then shown in Fig.~\ref{fig:np-mub}.
Then, at any fixed value of the correlation length $\xi$ the $n_p$-dependence that enters the expressions (\ref{eq:om}) and (\ref{eq:om2}) for $\omega_{4p}$ and $\omega_{4(p-\overline{p})}$ is
\begin{equation}
\left(\frac{n_p}{n_0}\right)^3 =\ 0.34,\ 0.77,\ 4.9,\ 31
\label{ProtonBottomline}
\end{equation}
and
\begin{equation}
\left( \frac{n_{p-\overline{p}}}{n_0} \right)^3 \left( \frac{   n_{p-\overline{p}}  } {  n_p+n_{\overline{p}}} \right)=
0.00072,\ 0.064,\ 3.4,\ 30 ,
\label{NetProtonBottomline}
\end{equation}
respectively, when evaluated at $\sqrt{s}=200,\,62,\,19$ and 7.7 GeV.  Since $n_{p-\bar p}$ is less than $n_p$ (and consequently less than $n_{p}+n_{{\bar p}}$) at all $\mu_B$ --- see Fig.~\ref{fig:np-mub} --- the critical contribution to $\omega_{4p}$ is greater than the critical contribution to $\omega_{4(p-\bar p)}$ at all $\mu_B$. The analogous argument applies  in comparing any $\omega_{i(p-\bar p)j\pi}$ to the corresponding $\omega_{ipj\pi}$. In all cases the suppression of the critical contribution to $\omega_{i(p - \bar p)}$ is most accute at small $\mu_B$, meaning 
at large $\sqrt{s}$.   We shall ignore $\omega_{i(p-\bar p)j\pi}$ in Section~\ref{section3}.


Second, we can ask which observable is most sensitive to critical fluctuations.  
For a given $\xi$, the critical contribution to $\omega_{ipj\pi}$ is largest when $r$ is largest, since as we see from (\ref{eq:om}) this gives $\omega$ the strongest $\xi$-dependence.
The experimental measurements reported in Ref.~\cite{Aggarwal:2010wy} demonstrate that $\kappa_{4(p-\bar p)}/\kappa_{2(p-\bar p)}$ can be measured with error bars that are much smaller than 1, 
and we expect that $\omega_{4p}$ and $\omega_{4\pi}$ can be measured with comparably small error bars.    The error bars on measurements of cumulants with $r>4$  will be larger, so until experimentalists demonstrate that they can be measured we have focussed on cumulants with $r\leq 4$.  For a given $\xi$ and $r$, the critical contribution to $\omega_{ipj\pi}$ is largest for $i=r$ if 
$\alpha_p^r/n_p > \alpha_\pi^r/n_\pi$ or for $i=0$ if 
$\alpha_p^r/n_p < \alpha_\pi^r/n_\pi$.

It is apparent from Figs.~\ref{fig:om4pa} and \ref{fig:om4pb} that 
$\omega_{4p}\gg \omega_{4\pi}$ at $\mu_B^c=400$~MeV with our 
benchmark values of $g_p$ and $g_\pi$, meaning that 
$\omega_{4p}$ is the normalized cumulant with $r=4$ that is most sensitive to critical fluctuations.  And, it is sensitive indeed:
we see from the plots in Fig.~\ref{fig:om4pa} that if $\xi$ reaches 2 fm, the critical contribution to $\omega_{4p}$ will be dramatic.  Correspondingly, if for example experimental measurements were to show that $\omega_{4p}-1 < 1$ at  some $\mu_B$ around 400 MeV, then $\xi<1$~fm at that $\mu_B$.
However, if $\mu_B^c$ is much less than 400 MeV and/or if $g_p/g_\pi$ is much smaller than with our benchmark values, then $\alpha_p^r/n_p$ could become less than $\alpha_\pi^r/n_\pi$, making $\omega_{4\pi}$ the best observable with which to find evidence for the presence of critical fluctuations.  
(With $g_p$ and $g_\pi$ set to their benchmark values, $\alpha_p^4/n_p = \alpha_\pi^4/n_\pi$ at $\mu_B\simeq 135$~MeV.)
Both $\omega_{4p}$ and $\omega_{4\pi}$  should be measured, and we shall see in Section III that if critical fluctuations are discovered it will be very important to have data on as many of the $\omega_{ipj\pi}$ as possible.


%
%
%
%
%
\section{Ratios of cumulants}
\label{section3}







In the previous Section, we presented numerical results for the contribution made by critical fluctuations to various cumulants of particle multiplicity distributions. 
In order to locate the critical point, experimental results on multiplicity cumulants will need to be compared to the theoretical predictions of the critical contribution to these cumulants. 
But, recall that we had to choose benchmark values for four parameters: $g_p$, $g_\pi$, $\widetilde{\lambda}_3$ and $\widetilde{\lambda}_4$. 
These parameters are not known reliably or accurately enough to permit a quantitative prediction for the effect of the critical point on any one of the cumulants we have described. 
The normalized cumulants depend on $\xi$, of course, but their dependence on the  four poorly known parameters would make it difficult to determine $\xi$ from data on any one of the cumulants, in isolation.   In this Section, we suppose that at some $\sqrt{s}$ there is experimental data showing several of the cumulants significantly exceeding their Poisson values.  We ask how ratios of cumulants can be used to extract information on $\xi$ and the values of the four parameters. And, we construct ratios of cumulants that are independent of $\xi$ and all the parameters, allowing us to make robust predictions for the contribution of critical fluctuations to these ratios.

The contributions of critical fluctuations to different correlators depend on different combinations of $\xi$ and the four parameters.
For example, 
\begin{eqnarray}
\kappa_{2p,\sigma} &\sim& V n_p^2 \,g_p^2 \,\xi^2,\nonumber \\
\kappa_{3p,\sigma} &\sim& V n_p^3\, g_p^3 \,\tilde{\lambda}_3 \,\xi^{9/2} ,\nonumber \\
\kappa_{4p,\sigma} &\sim& V n_p^4\, g_p^4 \,\tilde{\lambda}'_4 \, \xi^7,
\end{eqnarray}
where $ \tilde{\lambda}'_4 \equiv 2\tilde{\lambda}_3^2-\tilde{\lambda}_4  $.
For the most general pion-proton cumulant,
\begin{equation}
\kappa_{ipj\pi,\sigma} \sim V n_p^i \, g_p^i \, g_\pi^j \, \tilde\lambda'_r\,\xi^{\frac{5}{2}r-3}\ ,
\end{equation}
with $r=i+j$ and with $\tilde\lambda'_r$ as defined in (\ref{eq:lambda_r}).
We have kept the $n_p$-dependence since it introduces significant $\mu_B$-dependence, but we have suppressed the $T$- and $n_\pi$-dependence.
In Table \ref{table:ratios} we present the parameter dependence of various cumulant ratios. Except for the first 3 entries, $N_\pi$, $N_p$ and $\kappa_{ipj\pi}$, the quantities we consider are all $V$-independent (i.e. intensive) by construction. (In constructing intensive ratios, we can always remove  $V$-dependence by dividing by $N_\pi$ to the appropriate power.)  Note that although we have not written the $\sigma$ subscripts in the table,  the table only describes the parameter-dependence of the contributions from critical fluctuations.  When the ratios in the table are constructed from data, the Poisson contribution must be subtracted from each measured $\kappa$ separately, before taking a ratio.  This means that this table will only be useful in the analysis of data at values of $\sqrt{s}$ at which several $\kappa$'s are different from their Poisson values by amounts large compared to the experimental statistical and systematic error bars.


\begin{table}  \caption{Parameter dependence of the contribution of critical fluctuations to various particle
    multiplicity cumulant ratios. We have subtracted the Poisson
    contribution from each cumulant before taking the ratio. The table
    shows the power at which the parameters enter in each case. We
    only considered cases with $\ipj\equiv i+j=2,\,3,\,4$. We defined
    $2\tilde{\lambda}_3^2-\tilde{\lambda}_4 \equiv
    \tilde{\lambda}'_4  $.}
\newcommand{\nn}{-}
\begin{tabular}
{|c||ccccccc|}	\hline
ratio &$V$& $n_p(\mu_B)$ &  $g_p$   &   $g_\pi$  & $\tilde{\lambda}_3$ & $ \tilde{\lambda}'_4  $  & $\xi$    \\

	\hline \hline

$N_\pi$ & 1 & \nn & \nn & \nn & \nn &  \nn & \nn \\
\hline
$N_p$ & 1 & 1 & \nn & \nn & \nn &  \nn & \nn \\
\hline
$\kappa_{ipj\pi}$& 1 & $i$ & $i$ & $j$ & $\delta_{\ipj, 3}$ & $\delta_{\ipj, 4}$ & $\frac{5}{2} \pipj -3$ \\
\hline\hline	
$\omega_{ipj\pi}$& \nn & $i-\frac{i}{\ipj}$ & $i$ & $j$ & $\delta_{\ipj, 3}$ & $\delta_{\ipj, 4}$ & $\frac{5}{2} \pipj -3$ \\
\hline
$\kappa_{ipj\pi} N_\pi^{i-1}/N_p^i $ & \nn & \nn & $i$ & $j$ & $\delta_{\ipj, 3}$ & $\delta_{\ipj, 4}$ & $\frac{5}{2} \pipj -3$ \\
\hline
\hline
$\kappa_{2p2\pi}N_\pi/\kappa_{4\pi}\kappa_{2p}$& \nn & \nn & \nn & $ -2  $ &\nn&\nn&$-2$\\  
\hline
$\kappa_{4p}N_\pi^2/\kappa_{4\pi} \kappa_{2p}^2$& \nn & \nn & \nn &$ -4 $& \nn & \nn &$-4$\\
\hline
\hline
$\kappa_{2p2\pi} N_p^2/\kappa_{4p} N_\pi^2$& \nn & \nn &$-2$&2& \nn & \nn & \nn \\
\hline
$\kappa_{3p1\pi} N_p/\kappa_{4p} N_\pi$& \nn & \nn &$-1$&1& \nn & \nn & \nn \\
\hline
\hline
$\kappa_{3p} N_p^{3/2}/\kappa_{2p}^{9/4}N_\pi^{1/4}$& \nn & \nn &$-3/2$& \nn &1& \nn & \nn \\
\hline\hline
$\kappa_{2p}\kappa_{4p}/\kappa_{3p}^2$& \nn & \nn & \nn & \nn &$-2$&1& \nn \\
\hline\hline
$\kappa_{3p}\kappa_{2\pi}^{3/2}/\kappa_{3\pi} \kappa_{2p}^{3/2}$& \nn & \nn & \nn & \nn & \nn & \nn & \nn \\
\hline
$\kappa_{4p}\kappa_{2\pi}^2/\kappa_{4\pi} \kappa_{2p}^2$& \nn & \nn & \nn &
\nn & \nn & \nn & \nn \\
\hline
$\kappa_{4p}^3\kappa_{3\pi}^4/\kappa_{4\pi}^3 \kappa_{3p}^4$& \nn & \nn & \nn &
\nn & \nn & \nn & \nn \\
\hline
$\kappa_{2p2\pi}^2/\kappa_{4\pi} \kappa_{4p}$& \nn & \nn & \nn &
\nn & \nn & \nn & \nn \\
\hline
$\kappa_{2p1\pi}^3/\kappa_{3p}^2 \kappa_{3\pi}$& \nn & \nn & \nn &
\nn & \nn & \nn & \nn \\
\hline
\end{tabular}  \label{table:ratios}
\end{table}


Looking at  Table~\ref{table:ratios}, one can see how to use 
cumulant ratios in order to constrain $\xi$ and  the four parameters. 
The correlation length $\xi$ and the four nonuniversal parameters always appear in certain
combinations in the multiplicity cumulants and it turns out that we can only
constrain four independent  combinations. We have constructed the table to highlight ratios that can be used to constrain one example of four such combinations, with each block delineated by double horizontal lines corresponding to constraining
\begin{enumerate}

\item $g_\pi\,\xi$ --- using, e.g., $\kappa_{2p2\pi}N_\pi/\kappa_{4\pi}\kappa_{2p}$ or 
$\kappa_{4p}N_\pi^2/\kappa_{4\pi} \kappa_{2p}^2$.\footnote{The ratio
$\omega_{2\pi}=\kappa_{2\pi} /N_\pi$ could also be used here. However, we have seen that the critical contribution to this quantity is
small and, given the multitude of alternative choices, we can afford
not to use this quadratic moment.}

\item $g_\pi/g_p$ --- using, e.g., $\kappa_{2p2\pi}
  N_p^2/\kappa_{4p} N_\pi^2$ or $\kappa_{3p1\pi} N_p/\kappa_{4p} N_\pi$.

\item $\tilde{\lambda}_3^2/g_p^{3}$ --- using, e.g., $\kappa_{3p} N_p^{3/2}/\kappa_{2p}^{9/4}N_\pi^{1/4}$.

\item $\tilde{\lambda}'_4/\tilde{\lambda}_3^2$ --- using, e.g., $\kappa_{2p}\kappa_{4p}/\kappa_{3p}^2$.

\end{enumerate}
Since four independent combinations of $\xi$ and the  four parameters can be 
constrained by data on these ratios, we could, for example,  use data to express $\xi$ and three
of the parameters in terms of the fourth, say $g_p$.


We can see from Table \ref{table:ratios} that there are also some
combinations (e.g., the last five entries in the table) that are
parameter-independent. The first two of these are in fact the ratios of
the skewness and kurtosis of protons to pions, 
where skewness and kurtosis are defined as usual as
\begin{equation}
\mathrm{skewness}=\frac{\kappa_3}{\kappa_2^{3/2}},\:\:\:\:\:\: \mathrm{kurtosis}=\frac{\kappa_4}{\kappa_2^{2}}.
\end{equation}
The next row in the table is the ratio of the two rows above it, giving a combination that has the virtue that it only involves 3rd and 4th cumulants, which is advantageous since the contribution of the critical fluctuations is larger at larger $r$.
The last two ratios in the table are quite different, as they involve mixed cumulants, but they too are parameter-independent.
So, the last five ratios in the table have no $\xi$-dependence, no dependence on the four poorly known parameters, and no $n_p$-dependence.  This means that, after we subtract the Poisson contribution to each of the cumulants involved, we can make a robust prediction for the ratios of the contributions of critical fluctuations.
We find that these five ratios are all precisely 1.


Now let us see how we can use these five ratios in order to locate the
critical point. Suppose that as you change the center of mass energy
$\sqrt s$ of the collisions there is a point where many cumulants exceed their Poisson values by statistically significant amounts.  
As we see from 
Figs. \ref{fig:om4pa} and \ref{fig:om4pb},
the qualitative signature of the critical point is peaks in the multiplicity
cumulants as a function of $\sqrt{s}$.   Suppose experimental evidence for such peaks begins to emerge.  The specific ratios of the heights of the peaks in 
Figs. \ref{fig:om4pa} and \ref{fig:om4pb} depended on the benchmark choices for parameters that we made in those figures. So, how do you check in a parameter-independent fashion whether the behavior seen in experimental data is consistent with the hypothesis that it is due to critical fluctuations? You first subtract the Poisson contributions,\footnote{This can be done by subtracting the values of the cumulants in a sample of ``mixed events,'' i.e. events constructed artificially from tracks drawn at random from many different events in order to remove all correlations.  In this way, in addition to subtracting Poisson fluctuations one will also subtract spurious experimental effects. The mixed event technique is widely used in the study of quadratic moments and it could be used here too, even though present data indicate that spurious experimental effects are quite small~\cite{Aggarwal:2010wy}.
} 
and then construct the last five ratios in Table I.  If the fluctuations seen in this hypothetical data are in fact due to the proximity of the critical point, all five of these ratios will be equal to 1, with no theoretical uncertainties arising from uncertainty in the values of the parameters.  This would be strong evidence indeed for the discovery of the QCD critical point.

\section{Discussion}
\label{discussion}







We have explored the effects of the long wavelength fluctuations that arise in heavy ion collisions that freezeout near the QCD critical point
on higher cumulants of particle multiplicities. 
The characteristic signature is the non-monotonic behavior of such observables as a function of the collision energy, as the freezeout point approaches and then passes the critical point in the QCD phase diagram.    In Section I we illustrated one possibility for how seven different cumulants (third and fourth cumulants of protons and pions plus three mixed cumulants) might behave as a function of $\mu_B$, the chemical potential at freezeout which is the quantity that a beam energy scan scans.  In Section II, after calculating 21 different cumulants as a function of parameters and as a function of the correlation length $\xi$ at freezeout, we determined that either $\omega_{4p}$ or $\omega_{4\pi}$ is the most sensitive to critical fluctuations, depending on the values of parameters and depending on the location of the critical point.  However, if critical fluctuations are discovered it will be important to have measured as many of the cumulant observables as possible. In Section III we constructed ratios of observables that will allow an overconstrained experimental determination of currently poorly known parameters. And, we constructed other ratios of observables that, if the measured cumulants are indeed dominated by critical fluctuations, are independent of all the parameters in our calculation and are independent of the value of the correlation length.  We are therefore able to make parameter-independent predictions for these ratios, predictions that we hope will some day make it possible to determine with confidence that observed fluctuations do indeed indicate proximity to the critical point.


There are several effects that require further investigation:
\begin{itemize}

\item
In our treatment of the pions we have assumed that all pions seen in the detector reflect the physics at the time of freezeout, but it is thought that roughly half of the detected pions come from the later decay of resonances~\cite{Stephanov:1999zu}.  
Let us look at how this affects our results. Consider the 
peaks in Figs.~\ref{fig:om4pa} and \ref{fig:om4pb}, in the vicinity of $\mu_B^c$ where freeze-out occurs closest to the critical point.  Because the cumulants (for example $\kappa_{4\pi}$) are extensive, our calculation of the normalized cumulants (for example 
$\omega_{4\pi}=\kappa_{4\pi}/\langle N_\pi\rangle$) would be correct if the experimentalists measuring 
$\omega_{4\pi}$ divide by the number of pions present at freezeout.   By dividing instead by the number of pions seen in the detector, the experimentalists will obtain a smaller $\omega_{4\pi}$ than in our calculation.  This is an effect that can be corrected for.  

\item
There are physical effects that can limit the upward fluctuation of $N_p$.  For example, if the proton number density becomes too large, it will not be a good approximation to treat the protons at the time of chemical freezeout as noninteracting.  In Appendix A we make a crude attempt to estimate the consequences of including such effects on the normalized cumulants.  It will be much easier to model the consequences of this effect with data that show evidence for critical fluctuations in hand, since such data itself will indicate whether upward fluctuations in $N_p$ are cutoff, and if so at what $N_p$.

\item
The fact that net baryon number is conserved will also limit the fluctuation in $N_p$. The 
magnitude of this effect
depends on the size of the acceptance window, and for noncritical (and
Gaussian) fluctuations has been studied 
in Refs.~\cite{Jeon:2000wg,Asakawa:2000wh,Shuryak:2000pd,Jeon:2003gk,Koch:2008ia}). It also
depends on the features of baryon number fluctuations {\em outside} the
acceptance window.  It may translate into a sharp cutoff on the upward
fluctuation of $N_p$ (as, e.g., proposed in a model study in Ref.~\cite{Schuster:2009jv}) or
the reduction in flucutations may be more smoothly distributed over a range of $N_p$. We defer investigation of this effect to future work.
Experimentalists will also be able to learn more about this and other effects by
studying the dependence of the normalized proton cumulants on the width of the rapidity acceptance window, once there is data showing evidence of critical fluctuations.

\item
We have focussed on fourth and lower order cumulants.  Our results show, though, that higher order cumulants depend on even higher powers of the correlation length $\xi$, making them even more sensitive to the proximity of the critical point.  However, the measurement of higher order cumulants  involve the subtraction of more and more terms, making it important to determine the precision with which they can be measured.  We have stopped at fourth order because current analyses show that these cumulants can be measured with small error bars.  If cumulants beyond fourth order are measured, it will be possible to construct further ratios of observables that overconstrain parameters or are independent of parameters.

\end{itemize}


%
%
%
%
%

\acknowledgments

We would like to thank Sourendu Gupta, Bedanga Mohanty, Nu Xu and Zhangbu Xu for very helpful discussions. 
This research was
supported in part by the DOE Office of Nuclear
Physics under grants \#DE-FG02-94ER40818 and  \#DE-FG0201ER41195.

\appendix

\section{Toy Model Probability Distribution}
\label{appendixToy}
In Section \ref{subsection2} we presented the calculation of (and results for) the second, third and fourth cumulants of the pion, proton and net proton multiplicity distributions.  We found, for example, that with our benchmark parameters $\omega_{4p}$ peaked at a value of around 400, while for a Poisson distribution $\omega_{4p}=1$.  This dramatic increase in $\omega_{4p}$ due to critical fluctuations with a reasonable value of $\xi=\xi_{\rm max}=2$~fm
raises the question of what the distribution whose moments we calculated looks like --- does it in any way look unreasonable or unphysical?  Although the results from Section \ref{subsection2} determine arbitrarily high cumulants of the proton multiplicity distribution, they do not allow us to determine the shape of the distribution itself.  In this Appendix, we provide
an example of a probability distribution ${\cal P}(N_p)$ which has values of $\omega_{ip}$ for $i=2$, 3 and 4 that are comparable to those we calculated in Section \ref{subsection2}.  This toy model distribution is somewhat, but not completely, {\it ad hoc}, since we shall construct it in a way that does reflect the origin of the critical contribution to the fluctuations.

Let us consider a free gas of particles of a given species that have a  mass $M(\sigma)$ which is a function of a background field $\sigma$. 
As an example, for protons we shall use $M(\sigma) = m_p+g_p \sigma$.   
And, let us assume that the $\sigma$ field fluctuates with a given probability 
distribution ${\cal P_\sigma}(\sigma)$.   The central simplification that we are making in constructing this toy model is that we are assuming that $\sigma$ is spatially homogeneous.  The field $\sigma$  fluctuates, but at any given time it is the same everywhere in space.  
Let us model the probability distribution for the number $N_p$ of particles with mass $M(\sigma)$ by considering the fluctuations of $N_p$ due to the fluctuations in $\sigma$. Integrating over the fluctuating $\sigma$ we obtain the probability distribution for $N_p$:
\begin{equation}
  \label{eq:P}
  {\cal P}(N_p) 
= 
\int d\sigma\,{\cal P}_{\sigma}(\sigma)\, P_{M(\sigma)}(N_p)
\end{equation}
where $P_{M}(N)$ is the probability distribution for a
particle with {\em fixed} mass $M$ which we choose to be Poisson:
\begin{equation}\label{eq:Poisson-N}
   P_{M}(N_p) = \frac{\bar N^{N_p}}{N_p!}e^{-\bar N},
\end{equation}
where $\bar N$ is the expectation (mean) value of $N_p$ for the
distribution $P_{M}(N_p)$. In thermal and chemical equilibrium,
\begin{equation}
  \label{eq:nbar}
  \bar N = V \int_\mathbf{\p} \frac{1}{\exp
  \left(\frac{\sqrt{\p^2+M(\sigma)^2}-\mu}{T}\right) \pm 1}
\end{equation}
where we choose the positive sign since protons are fermions.  $ \bar N$ depends on $M$ and,
therefore, on $\sigma$.

The probability distribution of $\sigma$ is determined by the effective
potential $\Omega(\sigma)$:
\begin{equation}
  \label{eq:OmegaAppDefn}
  {\cal P}_\sigma(\sigma) \sim \exp\left(-V\,\frac{\Omega(\sigma)}{T}\right),
\end{equation}
where the effective potential can be written as
\begin{equation}
  \label{eq:OmegaApp}
  \Omega(\sigma) = \frac{m^2}{2}\sigma^2 
+ \frac{\lambda_3}{3}\sigma^3
+\frac{\lambda_4}{4}\sigma^4 + \ldots\, ,
\end{equation}
namely (\ref{eq:Omega}) without the spatial gradients.
Eqs.~(\ref{eq:P})-(\ref{eq:OmegaApp}) define the probability
distribution for the particle number, which will depend, among other
things, on the correlation length $\xi\equiv m_\sigma^{-1}$.   Note that since the volume $V$ in the model corresponds to the volume within which the $\sigma$ field is homogenous we should think of $V$ as a parameter in the toy model just as $\xi$ is.
The model
treats only the zero-momentum mode  $\sigma=\int_{\bm x}\sigma(\bm x)/V$ of
the critical field, ignoring all other modes, i.e., the space
variation of the field $\sigma(\bm x)$. This means, in
particular, that it ignores the fact that the correlations are
exponentially small between regions of space separated by distances
further than $\xi$. For this reason we should not choose a value of $V^{1/3}$ that is very much larger than $\xi$.

\begin{figure}
  \centering
  \includegraphics[width=\linewidth]{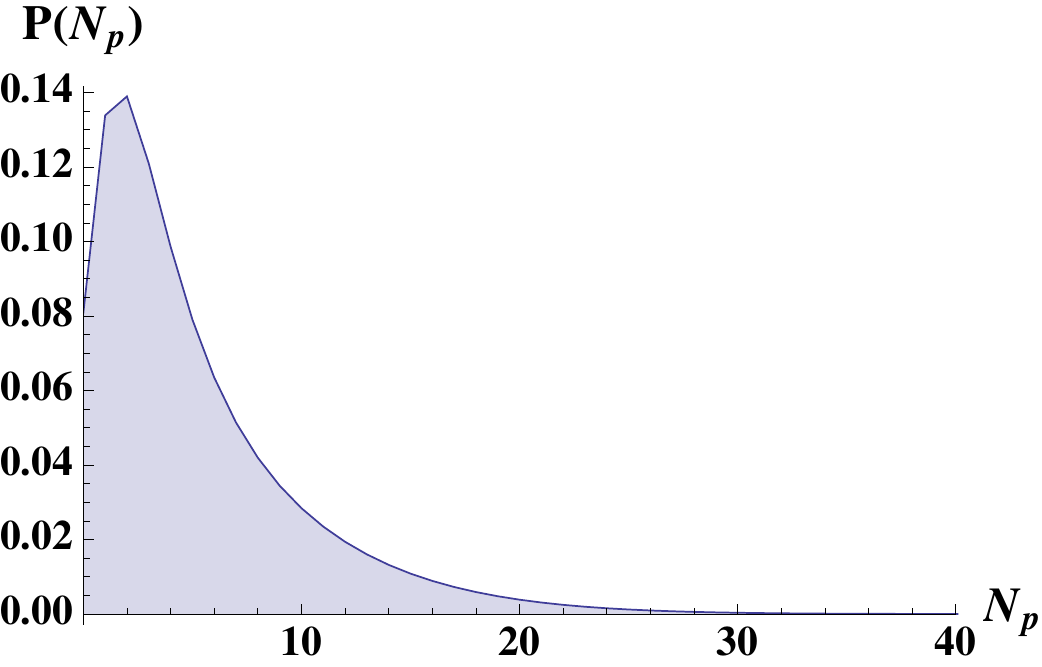}
  \caption{An example of a distribution with $\omega_{4p}^{\rm model}\approx 400$. The construction of the model distribution is described in the text, as are the values of its first few cumulants. $N_p$ is the number of protons in a volume $V=(5\ {\rm fm})^3$ in the toy model distribution. Other parameter choices are described in the text.
  }
   \label{fig:example-dist}
\end{figure}

As an example, in Fig.~\ref{fig:example-dist} we plot the toy model probability distribution for the number of protons, ${\cal P}(N_p)$ of (\ref{eq:P}), with $\xi=2$~fm, $\tilde\lambda_3$ and $\tilde\lambda_4$ taking their benchmark values, $\mu_B=400$~MeV, $V=(5 \:\rm{fm})^3$ and $g_p=6.185$.
We can then evaluate the mean and cumulants of this toy model probability distribution, and we find
\begin{eqnarray}
\langle N_p \rangle^{\rm model}&=&5.2, \nonumber\\
\omega_{2p}^{\rm model}&=&4.5, \nonumber\\  
\omega_{3p}^{\rm model}&=&37, \nonumber\\ 
\omega_{4p}^{\rm model}&=&405.
\end{eqnarray}
We chose all parameters in the toy model at their benchmark values with the exception of $g_p$, whose benchmark value is 7.
We chose $g_p=6.185$ in the toy model in order to get a probability distribution whose fourth cumulant is similar to that we calculated in Section \ref{subsection2}.  In our full calculation of Section \ref{subsection2}, with $\xi=2$~fm, $\mu_B=400$~MeV, and all parameters at their benchmark values including in particular $g_p=7$ we obtain
\begin{eqnarray}
\langle N_p \rangle&=&3.0, \nonumber\\
\omega_{2p}&=&4.2, \nonumber\\ 
 \omega_{3p}&=&30, 
\nonumber\\ 
\omega_{4p}&=&405\ ,
\end{eqnarray}
where we have quoted $\langle N_p \rangle = V n_p$ for $V=(5~{\rm fm})^3$.  (The $\omega$'s calculated in Section II are intensive, meaning that they are the same for any choice of $V$.)
We see that the distribution in Fig.~\ref{fig:example-dist} has cumulants that are similar to those we calculated in Section \ref{subsection2}, including in particular having as dramatically
 large a value of $\omega_{4p}$.  We see from the figure that an $\omega_{4p}$ that is $\sim 400$ 
 times larger than the Poisson value does not indicate an unusual looking distribution.  We can also note that the large positive 4th cumulant is a consequence of the skewness of the distribution.  If the distribution were symmetric, the large positive 4th cumulant would indicate a highly peaked distribution, but not so here.


We can also use our toy model to make a crude estimate of how our results for the cumulants would be affected by an upper cutoff on $N_p$.   In a heavy ion collision, $N_p$ (say in one unit of rapidity) cannot fluctuate to arbitrarily large values.    We can implement this in the toy model by putting an upper cutoff on $N_p$.  Lets assume that the neutrons are fluctuating with the protons, as is in fact expected~\cite{Hatta:2003wn}.  
It seems clear that nucleon-nucleon repulsion (that we have not taken into consideration) would cut off upward fluctuations in $N_p$ somewhere below those that correspond to nucleon densities of 1/fm$^3$, meaning $\sim 60$ protons per $(5~{\rm fm})^3$ volume.  To get a sense of the size of these effects, we tried cutting off the distribution in Fig.~\ref{fig:example-dist} at $N_p=30$.  We find
\begin{eqnarray}
\langle N_p \rangle^{\rm cutoff\ model}&=& 5.1\ , \nonumber\\
\omega_{2p}^{\rm cutoff\ model}&=&4.4\ , \nonumber\\  
\omega_{3p}^{\rm cutoff\ model}&=&33\ , 
\nonumber\\ 
\omega_{4p}^{\rm cutoff\ model}&=&289\ .
\end{eqnarray}
We see that a cutoff like this has little effect on the 2nd and 3rd cumulants, but it does reduce $\omega_{4p}$ by 28\%.  (See also Ref.~\cite{Schuster:2009jv} for a study of the effects of introducing a cutoff at large $N_p$ in the absence of critical fluctuations.)

\section{Mean Transverse Momentum Fluctuations}
\label{appendixpT}

The correlators found in section \ref{subsection1} can also be used to estimate the effect of the long wavelength fluctuations in the vicinity of the 
critical point on higher cumulants of the mean transverse momentum $p_T$. For example, the cubic cumulant of the mean $p_T$ distribution around the all event mean $\overline{p_T}$, namely $\kappa_3(\delta p_T)$, is given by
%
\begin{align} \label{appeq1}
\kappa_3 (&\delta p_T) \equiv \langle \langle (p_T-\overline{p_T})^3 \rangle \rangle 
\nonumber\\
&= 
\frac{1}{\left( \int_\mathbf{\p} \langle n_\mathbf{\p} \rangle \right)^3}
\int_\mathbf{\p_1}\int_\mathbf{ \p_2}\int_\mathbf{ \p_3} ([\mathbf{\p_1}]_T-\overline{p_T})  ([\mathbf{\p_2}]_T-\overline{p_T}) \notag\\
&\qquad\qquad\quad\times([\mathbf{\p_3}]_T-\overline{p_T})\langle \langle \delta n_\mathbf{\p_1} \delta n_\mathbf{\p_2} \delta n_\mathbf{\p_3} \rangle \rangle \ ,
\end{align}
%
and similarly for $\kappa_4(\delta p_T)$. We can normalize $\kappa_k(\delta p_T)$ by defining a dimensionless and intensive variable $F_k$:
\begin{equation} \label{appeq2}
F_k \equiv \frac{\langle N\rangle^{k-1} \kappa_k(\delta p_T)}{v_{\rm inc}^k(p_T)},
\end{equation}
where $\langle N\rangle$ is the total particle multiplicity and $v_{\rm inc}^2(p_T)$ is the variance of the inclusive (single-particle) $p_T$-distribution, defined as
\begin{equation}
v_{\rm inc}^2(p_T) =
\frac{1}{\int_\mathbf{\p} \langle n_\mathbf{\p} \rangle }
\int_\mathbf{\p}  \left(\mathbf{\p}_T- \overline{p_T} \right)^2 \langle n_\mathbf{\p} \rangle \ .
\end{equation}
Upon evaluating (\ref{appeq2}) using the correlators given in section \ref{subsection1}, we obtain the critical contribution to $F_k$. For pions with $\mu_\pi=0$ at $T=120$ MeV and $\xi=2$ fm, we find
\begin{equation}
F^\sigma_3=-0.0131 \:\:\:\mathrm{and} \:\:\:F^\sigma_4=0.0177.
\end{equation}
In addition to the critical point contribution, expression (\ref{appeq2}) receives contributions from Poisson statistics, Bose-Einstein enhancement, resonances, effects of radial flow, etc. It was shown in \cite{Stephanov:1999zu} that the effects of resonances and radial flow are very small and hence we will ignore them. Here we compare the critical point contribution to that coming from Bose-Einstein enhancement. The 3- and 4-particle correlators for an ideal Bose gas are given by
\begin{eqnarray} \label{appeq3}
\langle \langle (\delta n_\mathbf{k})^3 \rangle \rangle_{\mathrm{BE}}&=&  \langle n_\mathbf{k}\rangle  (\langle n_\mathbf{k}\rangle  +1)(2 \langle n_\mathbf{k}\rangle +1), \\ \label{appeq4}
\langle \langle (\delta n_\mathbf{k})^4 \rangle \rangle_{\mathrm{BE}}&=& \langle n_\mathbf{k}\rangle ( \langle n_\mathbf{k}\rangle  +1)(1+6  \langle n_\mathbf{k}\rangle ( \langle n_\mathbf{k}\rangle +1)),\qquad
\end{eqnarray}
where here by $\langle n_\mathbf{k}\rangle$ we mean the mean occupation number for an ideal Bose gas. Evaluating (\ref{appeq2}) using the above correlators will give us the Bose-Einstein and the Poisson contribution to $F_k$. In order to isolate the Bose-Einstein effect we subtract the Poisson contribution $\langle n_\mathbf{k} \rangle $ from $\kappa_k$ and then evaluate $F_k$. Using the same parameters as above we find
\begin{equation}
F^{\mathrm{BE}}_3=-0.2480\:\:\: \mathrm{and}\:\:\: F^\mathrm{BE}_4=0.9388.
\end{equation}
We see that the contribution of critical fluctuations is smaller than that due to  Bose-Einstein effects. We conclude that it would be very difficult to use higher cumulants of the mean $p_T$ distribution in order to search for the critical point. Furthermore, as kinetic freeze-out (where particle momenta freeze) occurs after chemical freeze-out (where particle numbers freeze), it is easier for $p_T$ fluctuations to get washed out (see, e.g. Ref.~\cite{Stephanov:2009ra}), making them even less favorable observables in searching for the critical point.


\end{document}